%% file: main.tex
\documentclass[a4paper,11pt]{article}
\pdfoutput=1 % if your are submitting a pdflatex (i.e. if you have
\usepackage{jheppub} % for details on the use of the package, please
\usepackage[T1]{fontenc}
\usepackage[utf8]{inputenc}

\include{./preambles/preambles}
\include{./preambles/math}

\usepackage[normalem]{ulem}
\newcommand\hl{\bgroup\markoverwith
    {\textcolor{yellow}{\rule[-.5ex]{.1pt}{2.5ex}}}\ULon}

\title{Semi-classical predictions of cosmological wave-packets from ridge-lines} % version-1

\author[a]{Chen Lan}
\author[b]{and Yi-Fan Wang}
\affiliation[a]{School of Physics, Nankai University,\\Weijin Road 94, Tianjin 300071, China}
\affiliation[b]{Institute for Theoretical Physics, University of Cologne,\\Zülpicher Str.\ 77a, 50937 Cologne, Germany}

\emailAdd{lanchen@nankai.edu.cn}
\emailAdd{yifan.wang@live.com}

\abstract{We introduce a concept
of ridge-lines to investigate the semi-classical prediction from wave-packets 
with arbitrary width in conventional quantum mechanics and
the Wheeler--DeWitt quantum cosmology. 
Two primary approaches are applied to the exact calculation of the
ridge-lines, namely the contour and the stream approach. 
Moreover, aspects of these are discussed and compared 
to other scenarios and approaches, i.e.\ the narrow WKB wave-packets and 
the first-derivative test.
As the main result, we show that the semi-classical predictions in toy models have more abundant solutions than in the classical theory, and most interestingly they may deviate from classical solutions due to the quantum corrections.}

\begin{document}

\maketitle

% \tableofcontents
 
%\newpage 

%1234567890123456789012345678901234567890123456789012345678901234567890123456789
\section{Introduction}
\label{sec:intro}
%1234567890123456789012345678901234567890123456789012345678901234567890123456789 

As the prevailing theory of gravitation, the general theory of relativity successfully describes classical gravitation, but has yet to be consistently quantised, despite the efforts of generations of physicists in over a hundred years.

One of the first attempts to quantise general relativity directly is the Wheeler–-DeWitt approach,
see e.g.\ \cite{Kiefer2012,Bojowald:2015iga}.
It begins with the Hamiltonian formulation of this theory by Arnowitt, Deser and Misner, and applies the quantisation scheme of Dirac, designed for constrained systems, including the Dirac spinors and the Maxwell theory, among others. This approach, also known as quantum geometrodynamics, is successful 
with the semi-classical methods
of Wentzel–-Kramers–-Brillouin (WKB) \cite{dabrowski2006} and Born-–Oppenheimer \cite{Kamenshchik:2020yvs}, and has been applied to quantum models of universes and black holes.

Unfortunately, because of the constrained nature of general relativity (from another perspective, its diffeomorphism invariance), its quantised version à la Dirac lacks many properties that are crucial in conventional quantum theory. 
Particularly, a positive-definite scalar product
of quantum states is difficult to define, rendering the non-existence of a Hilbert space, and of the analysis of self-adjoint operators.
As a consequence, one cannot use the usual way to make predictions as in 
conventional quantum mechanics. This includes, on the one hand, interpreting 
the inner product as a probability amplitude; on the other hand, analysing 
self-adjoint operators and studying their spectra.

Quantum cosmology is an application of quantum geometrodynamics to the universe as a whole, see e.g.\ \cite[ch.\ 8]{Kiefer2012},
where the Wheeler--DeWitt equation plays the central role.
The emergence of classical trajectories 
can be realised if the forms of \emph{wave functions} are similar to the 
\enquote{coherent states},
which strongly peak about a single classical trajectory \cite{Halliwell2009}.
However, such an analogue of \enquote{coherent state}
can only be achieved for simple examples.

In contrast, the emergence of classical trajectories from \emph{wave-packets} is 
relatively natural, where the 
wave-packets of universe are constructed by the superposition of wave functions, 
and follow the classical trajectories in configuration space, 
when their width becomes narrow \cite{Kazama:1985cg,kiefer1988,Kiefer1990}.
This corresponds to the principle of constructive interference.  
Nevertheless, the correspondence between classical and quantum theories can only be implemented 
with the narrowness condition.
In this work, we try to address these problems by porting conventional tools in physics and mathematics to this context, aiming to derive the classical prediction from wave-packet with arbitrary width.

The outline of topics in each section is as follows.
In section \ref{sec:prototype}, 
we summarise 
previous results of a two-dimensional minisuperspace model \cite{Andrianov2018wdx}, 
which can be solved exactly and will be used as the basics to investigate the classical prediction from the
corresponding wave-packets in this paper. 
Next, under the WKB approximation, we show that a \emph{narrow} Gaussian wave-packet has \enquote{maxima} on the classical trajectory 
and can be compared to the one derived by the method of ridge-lines discussed later,
that works for wave-packets with \emph{arbitrary} width.
In section \ref{sec:sta-wp}, 
we construct a framework of stationary wave-packets, that makes sense for both the minisuperspace Wheeler–DeWitt equation and the stationary Schr\"odinger equation.  The framework is then tested by the model of a two-dimensional hydrogen atom.
In section \ref{sec:ridge-line}, 
the concept of ridge-lines is 
introduced,
and an intuitive approach, the first-derivative test, is applied to derive the ridge-lines from wave-packets. 
The deviation from classical theory emerges apparently near the turning point, which is interpreted as a quantum correction.  
In sections \ref{sec:hr-pg} and \ref{sec:hr-sg}, we discuss two 
further approaches to find the ridge-lines as classical predictions from wave-packets with \emph{arbitrary} width, one is the contour approach, the other one is the stream approach. 
We provide exact mathematical descriptions of ridge-lines, which were historically developed for Riemannian geometry with a Euclidean metric signature. Then we try to generalise these descriptions to the pseudo-Riemannian geometry with a Lorentzian metric signature, which is the usual case of minisuperspaces. 
After that, we apply both approaches
in various examples, and discuss their advantages as well as deficits.
The section \ref{sec:hr-dif} includes a discussion of the relation between these two approaches.
Finally, we give a summary and conclusion in
section \ref{sec:sum-out},  as well as proposals of prospective physical applications.
The section \ref{app:prototype-wkb} 
collects the WKB approach used in section \ref{ssec:packet-gaussian}.

%1234567890123456789012345678901234567890123456789012345678901234567890123456789
\section{\boldmath A two-dimensional minisuperspace model}
\label{sec:prototype}
%1234567890123456789012345678901234567890123456789012345678901234567890123456789

In this section \ref{sec:prototype}, we study a prototype minisuperspace model that 
traces back to \cite{Andrianov2015,Andrianov2016,Andrianov2018wdx}, which 
is described by the 
minisuperspace action
\begin{subequations}
\begin{align}
S &= \mathrm{Vol}_3 \int \dd t\, \rfun{M}{t} \cbr{
	\mitsanss\rbr{-\frac{3}{\varkappa}\frac{\dot{\gamma}^2}{\rfun{M}{t}^2} +
	\frac{\mitsansl}{2} \frac{\dot{\chi}^2}{\rfun{M}{t}^2}} - V \ee^{g \chi}}
\label{eq:prototype-10} \\
&\eqqcolon \int \dd t\, 
	\cbr{\frac{1}{2\rfun{M}{t}}\mscrG_{IJ} q^I q^J -
		\rfun{M}{t}\rfun{\mscrV}{q}}\,,
\label{eq:prototype-11}		
\end{align}
\end{subequations}
where $\mitsanss^2 = \mitsansl^2 = \mitsansv^2 = 1$ are signs, $\mitsansv
\coloneqq \sgn V$, $g > 0$ is a coupling factor; $\mscrG_{IJ}$'s are the
components of the inverse minisuperspace DeWitt metric \cite{Kiefer2012},
$\mscrV$ the potential, and 
$q^I$ denotes collectively the minisuperspace 
variables 
$\cbr{\gamma, \chi}$
in configuration space. One sees that $M$ corresponds to a
lapse function and has no dynamics,
whereas $\gamma$ and $\chi$ are the dynamic variables.

This prototype model contains several homogeneous cosmological models as its 
special cases, including the closed Friedmann--Lema\^{\i}tre model with a
free scalar field \cite[sec.\ 8.1.2]{Kiefer2012}, the flat 
Friedmann--Lema\^{\i}tre model with a 
Liouville scalar field \cite{Andrianov2018wdx}, and the vacuum 
Kantowski--Sachs model. Moreover, it is exactly
solvable at both the classical and the quantum levels,
%(see table \ref{tab:bessel-10}), 
which facilitates the further study of the model.

At the classical level, the trajectory in the minisuperspace spanned by
$\rbr{\gamma, \chi}$ has %the form
a uniform representation
\begin{align}
	\ee^{g \chi}\, \rfun{\mathrm{trig}}%
	{\sqrt{\tfrac{3}{2\varkappa}}\,	g\rbr{\gamma - \gamma_0}}^2
	&= 
	\frac{\varkappa p_{\gamma}^2}{12 \mathrm{Vol}_3^2 \vbr{V}}\,,
	\label{eq:cla-traj-10}
\end{align}
where $\gamma_0$ is a constant, $\mathrm{trig}$ is a trigonometric function 
which depends on the four  
possible signs
$\rbr{\mitsansl, \mitsanss \mitsansv}$, see table \ref{tab:trig-10}.
\begin{table}
\begin{center}
\begin{tabular}{l|c}
\toprule
$\rbr{\mitsansl, \mitsanss \mitsansv}$ &
$\rfun{\mathrm{trig}}%
	{\sqrt{\tfrac{3}{2\varkappa}}g\rbr{\gamma - \gamma_0}}^2$ \\
\midrule
$\rbr{-, -}$ & 
$-\rfun{\sin}{\sqrt{\tfrac{3}{2\varkappa}}g\rbr{\gamma - \gamma_0}}^2$ \\	
$\rbr{-, +}$ &
$\rfun{\sin}{\sqrt{\tfrac{3}{2\varkappa}}g\rbr{\gamma - \gamma_0}}^2$ \\	
$\rbr{+, -}$ &
$\rfun{\sinh}{\sqrt{\tfrac{3}{2\varkappa}}g\rbr{\gamma - \gamma_0}}^2$ \\
$\rbr{+, +}$ & 
$\rfun{\cosh}{\sqrt{\tfrac{3}{2\varkappa}}g\rbr{\gamma - \gamma_0}}^2$ \\
\bottomrule
\end{tabular}
\end{center}
\caption[\boldmath Four cases of $\mathrm{trig}$ in eq.\ \eqref{eq:cla-traj-10}]%
	{Four cases of the trigonometrical function in
	eq.\ \eqref{eq:cla-traj-10}. The first case $\rbr{-,-}$ does not leave a real
	and physical trajectory for $\rbr{\gamma, \chi}$; $\rbr{-,+}$ gives %many
	 infinitely many isolated trajectories due to the periodicity of the sine function, $\rbr{+, -}$ gives two, and $\rbr{+, +}$ gives one.}
\label{tab:trig-10}
\end{table}

At the quantum level, 
the dynamics of cosmology is governed by 
the Wheeler--DeWitt equation \cite{DeWitt1967}
\begin{subequations}
\begin{align}
   0 &= \rfun{H_\perp}{\gamma, \chi, \frac{\hslash}{\ii}\partial_\gamma,
   \frac{\hslash}{\ii}\partial_\chi,
   }\rfun{\psi}{\gamma, \chi}\label{eq:MWDW-10}\\
   &=\sbr{\mitsanss\frac{\hslash^2}{\mathrm{Vol}_3}\rbr{
		+ \frac{\varkappa}{12}\partial_\gamma^2 - 
			\frac{\mitsansl}{2}\partial_\chi^2}
	+ \mathrm{Vol}_3 V \ee^{g \chi}} \rfun{\psi}{\gamma, \chi}\,,
\label{eq:prototype-wdw-10}
\end{align}
\end{subequations}
which provides a naive solution
\begin{align}
	\psi &\propto \ee^{\frac{\ii}{\hslash} p_{\gamma} \gamma}
		\rfun{\mathrm{Bessel}_{\vbr{\nu}}}{x}\,,
	\qquad \text{where}
	\label{eq:prototype-wdw-sol-10} \\
% \end{align}
% where
% \begin{align}
	\nu &\coloneqq \frac{1}{\hslash g}
		\sqrt{\frac{2\varkappa}{3}} p_{\gamma}\,,
	\qquad
	x \coloneqq 2\sqrt{2} \frac{\mathrm{Vol}_3 \sqrt{\vbr{V}}}{\hslash g}
		\ee^{g \chi/2}\,,
\label{eq:transform-20}
\end{align}
and $\rfun{\mathrm{Bessel}_\nu}{x}$ is %a 
the Bessel-type function of order $\nu$, the
type of which depends on the signs 
$\rbr{\mitsansl, \mitsanss \mitsansv}$, see table \ref{tab:bessel-10}, where
$\rfun{F_{\ii \nu}}{x}$ and $\rfun{G_{\ii \nu}}{x}$ are the unmodified Bessel 
functions adapted to purely imaginary orders, defined in \cite{dunster1990}.

%%%%%%%%%%%%%%%%%%%%%%%%%%%%%%%%%%%
\iffalse
%%%%%%%%%%%%%%%%%%%%%%%%%%%%%%%%%%
\begin{subequations}
\begin{align}
\rfun{F_{\nu}}{x} &\coloneqq \frac{1}{2} \cbr{
	\ee^{+\nu\pp\ii/2} \rfun{H^{(1)}_{\nu}}{x} + 
	\ee^{-\nu\pp\ii/2} \rfun{H^{(2)}_{\nu}}{x}
	}
	\\
	&\equiv \frac{1}{2} \rfun{\sec}{\frac{\nu\pp}{2}} \cbr{
		\rfun{J_{+\nu}}{x} + \rfun{J_{-\nu}}{x}}\,;
	\\
\rfun{G_{\nu}}{x} &\coloneqq \frac{1}{2\ii} \cbr{
	\ee^{+\nu\pp\ii/2} \rfun{H^{(1)}_{\nu}}{x} - 
	\ee^{-\nu\pp\ii/2} \rfun{H^{(2)}_{\nu}}{x}
	}
	\\
	&\equiv \frac{1}{2} \rfun{\csc}{\frac{\nu\pp}{2}} \cbr{
		\rfun{J_{+\nu}}{x} - \rfun{J_{-\nu}}{x}}\,.
\end{align}
\end{subequations}
%%%%%%%%%%%%%%%%%%%%%%%%%%%%%%%%%%%%
\fi
%%%%%%%%%%%%%%%%%%%%%%%%%%%%%%%%%%

\begin{table}
\begin{center}
\begin{tabular}{l|c}
\toprule
$\rbr{\mitsansl, \mitsanss \mitsansv}$ & 
$\rfun{\mathrm{Bessel}_{\nu}}{x}$
\\
\midrule
$\rbr{-, -}$ &
	$\textcolor{gray}{c_1 \rfun{K_{\nu}}{x} + c_2 \rfun{I_{\nu}}{x}}$ \\
$\rbr{-, +}$ &
	$c_1 \rfun{J_{\nu}}{x} + \,\textcolor{gray}{c_2 \rfun{Y_{\nu}}{x}}$ \\
$\rbr{+, -}$ &
	$c_1 \rfun{F_{\ii \nu}}{x} + c_2 \rfun{G_{\ii \nu}}{x}$\\
$\rbr{+, +}$ &
	$c_1 \rfun{K_{\ii \nu}}{x} + \,\textcolor{gray}{c_2 \rfun{I_{\ii \nu}}{x}}$\\
\bottomrule
\end{tabular}
\end{center}
\caption[Four cases of $\rfun{\mathrm{Bessel}_\nu}{x}$ in
	eq.\ \eqref{eq:prototype-wdw-sol-10}]%
	{Four cases of the Bessel function in eq.\ \eqref{eq:prototype-wdw-sol-10}. %that
%	solves eq.\ \eqref{eq:prototype-wdw-25}.
Branches that diverge at the infinite
	boundary are in \textcolor{gray}{grey}, which are to be dropped.
	%, see section \ref{ssec:bound-inf}.
	The remaining branches are all real and have no imaginary part.
	\label{tab:bessel-10}}
	%, see fig.\ \ref{fig:wav-fun-10}.}
\end{table}

%%%%%%%%%%%%%%%%%%%%%%%%%%%%%%%%%%%%%%%%%%%%
\iffalse
%%%%%%%%%%%%%%%%%%%%%%%%%%%%%%%%%%%%%%%%%%%
In eq.\ \eqref{eq:prototype-wdw-10}, imposing the self-adjointness
[\citealp[sec.\ II.9]{von_Neumann_1996};
\citealp[ch.\ VIII]{ReedVol1};
\citealp[sec.\ X.1]{ReedVol2}]
of the differential operation in $\cbr{\ldots}$ enables the
discussion of orthonormality \cites[sec.\ 8.5.3]{Gitman2012}.
The operation for the 
$\rbr{\mitsansl, \mitsanss \mitsansv} = \rbr{+, +}$-branch with 
$\rfun{K_{\ii \nu}}{x}$ is essentially self-adjoint, with a 
$\delta$-normalisation factor
\cite{Yakubovich2006,Passian2009,Szmytkowski2010}
\begin{align}
	N_{K_, \nu}^{-2} = \frac{\pp^2 }{k \nu\, \rfun{\sinh}{\pp\nu}}\,.
	\label{eq:norm-besselK-10}
\end{align}

%%%%%%%%%%%%%%%%%%%%%%%%%%%%%%%%%%%%%%%%%%%%
\fi
%%%%%%%%%%%%%%%%%%%%%%%%%%%%%%%%%%%%%%%%%%%

The $\rbr{\mitsansl, \mitsanss \mitsansv} = \rbr{-, +}$- and 
$\rbr{+, -}$-branches are \emph{not} essentially self-adjoint,
which was discovered in \cite{Andrianov2018wdx}; a family of
self-adjoint extensions is characterised by a number $a \in [0, 2)$. For 
$\rbr{+, -}$, the spectrum is continuous, and the
orthonormal
eigenfunction corresponding to $\nu$ is
\begin{align}
\rfun{\varXi_\nu^{(a)}}{y} = N_{\varXi, \nu}^{(a)} \rbr{
	\rfun{F_{\ii \nu}}{x}\cos\frac{\pp a}{2} +
	\rfun{G_{\ii \nu}}{x}\sin\frac{\pp a}{2}}\,,
\end{align}
where $N_{\varXi, \nu}$ is the $\delta$-normalisation factor \cite{Andrianov2018wdx}. 
%%%%%%%%%%%%%%%%%%%%%%%%%%%%%%%%%%%%%%%%%%%%
\iffalse
%%%%%%%%%%%%%%%%%%%%%%%%%%%%%%%%%%%%%%%%%%%
Adapting the method in \cite{Szmytkowski2010}, one can derive
\begin{subequations}
\begin{align}
&\int_{0}^{+\infty} \rfun{F_{\ii \nu_1}}{x} \rfun{F_{\ii \nu_2}}{x}
	\frac{\dd x}{x}
	= \frac{\tanh \frac{\pp \nu_1}{2}}{\nu_1} \rfun{\delta}{\nu_1-\nu_2}\,,
\\
&\int_{0}^{+\infty} \rfun{G_{\ii \nu_1}}{x} \rfun{G_{\ii \nu_2}}{x}
\frac{\dd x}{x}
= \frac{\coth \frac{\pp \nu_1}{2}}{\nu_1} \rfun{\delta}{\nu_1-\nu_2}\,,
\\
&\int_{0}^{+\infty} \sbr{\rfun{F_{\ii \nu_1}}{x} \rfun{G_{\ii \nu_2}}{x} +
\rfun{F_{\ii \nu_2}}{x} \rfun{G_{\ii \nu_1}}{x}}
\frac{\dd x}{x}
= 0\,;
\end{align}
therefore
% 我算的归一化在这里
\begin{align}
\rbr{N_{\varXi, \nu}^{(a)}}^{-2} = \frac{2}{k \nu}\rbr{
	\tanh\frac{\pp \nu}{2} \cos^2 \frac{\pp a}{2} +
	\coth\frac{\pp \nu}{2} \sin^2 \frac{\pp a}{2}}\,.
\end{align}
\end{subequations}
%%%%%%%%%%%%%%%%%%%%%%%%%%%%%%%%%%%%%%%%%%%%
\fi
%%%%%%%%%%%%%%%%%%%%%%%%%%%%%%%%%%%%%%%%%%%
For $\rbr{-, +}$, the spectrum is \emph{discrete} with
\begin{subequations}
\begin{align}
	\nu = 2n+a\,,\qquad n \in \BbbN\,,
\end{align}
and the corresponding orthonormal eigenfunctions read
\begin{align}
	\rfun{\varPhi_n^{(a)}}{y} = N_{J, n} \rfun{J_{2n+a}}{x}\,,
	\\
	\rbr{N_{J, n}^{(a)}}^{-2} = \frac{1}{k\rbr{2n+a}}\,.
\end{align}
\end{subequations}
These motivate the study of the minisuperspace model due to the potential of integrability.

%123456789012345678901234567890123456789
\subsection{An exact wave-packet}
\label{ssec:lwp}
%1234567890123456789012345678901234567890123456789012345678901234567890123456789

Like the stationary Schr\"odinger equation in conventional quantum mechanics,
the Wheeler--DeWitt equation is also a linear differential equation.
For a family of mode functions $\cbr{\psi_{\nu}}$, which are complete integrals 
of the Wheeler--DeWitt equation, one could therefore choose an
amplitude $\rfun{\mscrA}{\nu}$ and construct a \emph{wave-packet}
\begin{align}
	\varPsi = \int \dd \nu\,\rfun{\mscrA}{\nu} \psi_{\nu}\,,
	\label{eq:wav-pac-def-10}
\end{align}
which is a general solution of the Wheeler--DeWitt equation, independent of any
interpretations. It is scarce that an exact expression of a wave-packet in
minisuperspace models can be found. In this section \ref{ssec:lwp} we will study such
a case.

\begin{figure}
\begin{center}
\begin{subfigure}{.49\textwidth}
	\begin{center}
		\includegraphics{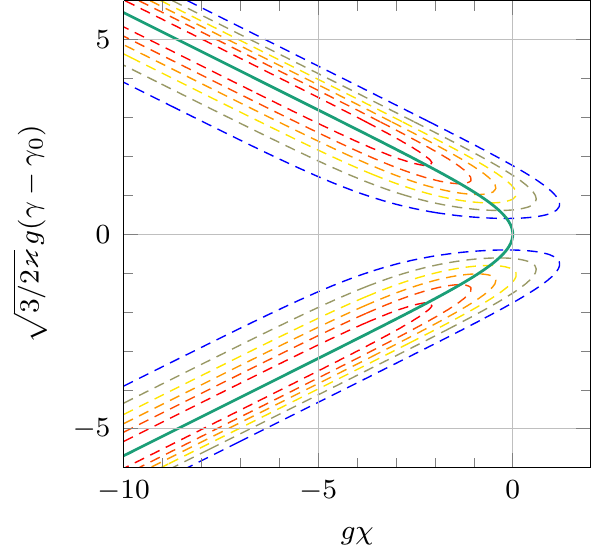}
	\end{center}
	\caption{$\mscrA_{\text{lin}}$
	\label{fig:linear-wp-10-lin}}
\end{subfigure}
\begin{subfigure}{.49\textwidth}
	\begin{center}
		\includegraphics{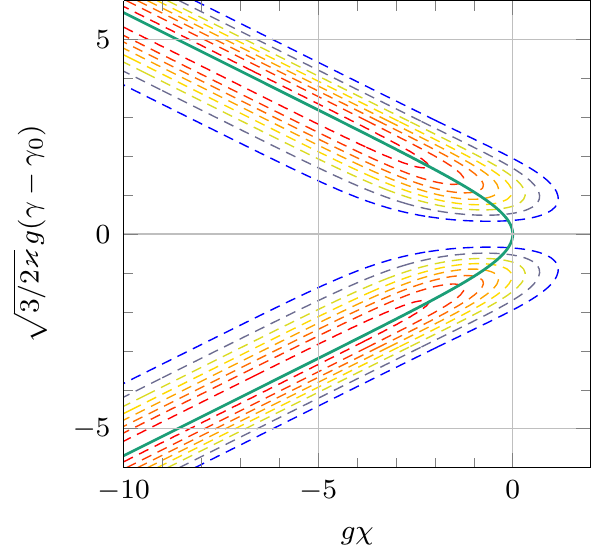}
	\end{center}
	\caption{$\mscrA_{\text{hfg}}$}
	\label{fig:linear-wp-10-gau}
\end{subfigure}
\end{center}
\caption[\enquote{Linear} wave-packets of the $\rbr{+, +}$-case]{
	Schr\"odinger profile $\vbr{\varPsi}^2$ of wave-packets of the 
	$\rbr{+, +}$-case of the prototype model, the
	mode-function of which is proportional to $\rfun{K_{\ii \nu}}{x}$.
	In fig.\ \ref{fig:linear-wp-10-lin}, the wave-packet is given by 
	eq.\ \eqref{eq:linear-pac-Psi}. The solid line is the classical trajectory with
	$\frac{\varkappa p_{\gamma}^2}{12 \mathrm{Vol}_3^2 \vbr{V}} = 1$, which 
	seems to lie \enquote{on the ridge} of the wave-packet. This will be 
	studied in section \ref{sec:ridge-line}. In fig.\ \ref{fig:linear-wp-10-gau}, a
	half-flipped Gaussian amplitude with respect to the normalised
	mode function, eq.\ \eqref{eq:amp-hfg-10}, is chosen.}
\label{fig:linear-wp-10}
\end{figure}

Making use of \cite[eq.\ (6.795.3)]{Gradshteyn2015a}, we have
\begin{align}
	\int_{-\infty}^{+\infty} \dd \nu\, \nu \ee^{\ii \nu y}
	\rfun{K_{\ii\vbr{\nu}}}{x} = \ii \pp x \ee^{-x \cosh y}\,\sinh y\,,
	\label{eq:6.795.3}
\end{align}
and are able to construct the exact wave-packet for the $\rbr{+,+}$-case of our
prototype model in table \ref{tab:bessel-10},
\begin{align}
\begin{split}
\rfun{\varPsi_\text{lin}}{\gamma, \chi} &\propto 
\ee^{\frac{g\chi}{2}}
	\sfun{\sinh}{\sqrt{\tfrac{3}{2 \varkappa }}\,g\rbr{\gamma -\gamma_0}}
\\
	&\quad\,\cdot
\cfun{\exp}{-\frac{2 \sqrt{2}\,\mathrm{Vol}_3 \sqrt{\vbr{V}}}{\hslash g}
		\ee^{\frac{g \chi }{2}} \sfun{\cosh}{\sqrt{\tfrac{3}{2 \varkappa }}\,
			g\rbr{\gamma -\gamma_0}}}\,,
\end{split}
\label{eq:linear-pac-Psi}
\end{align}
with an amplitude that \enquote{seems to be}
$\rfun{\mscrA_{\text{lin}}}{\nu} \propto p_{\gamma}\propto \nu$ 
(c.f.\ eq.\ \eqref{eq:transform-20}). This is a
typical profile of the norm square $\vbr{\varPsi}^2$ of a wave-packet in
Wheeler--DeWitt quantum cosmology, which forms a tube around some classical
trajectory in the asymptotic region, see fig.\ \ref{fig:linear-wp-10-lin}.

One may wonder how an amplitude that is proportional to the \enquote{wave 
number} $\nu$ can lead to a smooth wave-packet that makes physical sense. For 
example, if one naively takes plane waves $\rbr{2\pp}^{-1/2}\ee^{\ii k x}$ and 
uses a linear amplitude, one finds
\begin{equation}
    \frac{1}{\sqrt{2\pp}}\int_{-\infty}^{+\infty} \dd k\,k \ee^{\ii k x} =
	-\ii \sqrt{2\pp}\,
	\rfun{\delta'}{x}
\end{equation}
which is $0$ for $x \neq 0$.

The doubts can be dispelled if one considers the \emph{Schr\"odinger
normalisation} of $\rfun{\ii \vbr{\nu}}{x}$, which is given in
\cite{Yakubovich2006,Passian2009,Szmytkowski2010}, leading to the true amplitude
\begin{align}
	\rfun{\mscrA_{\text{lin}}}{\nu} \propto \frac{\nu}{N_{K, \nu}} \propto 
		\sqrt{\frac{\nu}{\rfun{\sinh}{\nu}}}\sgn\nu\,,
\end{align}
where we have used a $\delta$-normalisation factor $N_{K_, \nu} = k \nu\, \rfun{\sinh}{\pp\nu}/\pp^2$ for $\rfun{K_{\ii\vbr{\nu}}}{x}$.
In turn, the normalisation condition for the amplitude
$\int_{-\infty}^{+\infty} \dd\nu\,\vbr{\rfun{\mscrA_{\text{lin}}}{\nu}} = 1$
gives
\begin{align}
	\rfun{\mscrA_{\text{lin}}}{\nu} =
	\sqrt{\frac{2\nu}{\rfun{\sinh}{\nu}}}\sgn\nu\,.
	\label{eq:amp-linear-10}
\end{align}

To understand more about $\mscrA_{\text{lin}}$, one can turn to the 
\emph{Gaussian amplitude} that is popular in the literature, and compare 
the former with a modified version of the latter, which is flipped
with respect to the $x$-axis for $\nu < 0$ and has the same second moment 
$\abr{\nu^2}$ as $\mscrA_{\text{lin}}$. The second moment for
the \enquote{linear} amplitude in eq.\ \eqref{eq:amp-linear-10} reads
\begin{align}
	\int_{-\infty}^{+\infty} \dd\nu\,
	\nu^2 \vbr{\rfun{\mscrA_{\text{lin}}}{\nu}}^2 = \frac{1}{2}.
\end{align}
One therefore uses the \emph{one-dimensional Gaussian distribution}
%(c.f.\ eq.\ \eqref{eq:packet-gaussian-50b})
\begin{align}
	\rfun{\mathrm{GD}_1}{0, \sigma^2 = \frac{1}{2}; \nu} =
		\pp^{-1/2} \ee^{-\nu^2}
\end{align}
and constructs the amplitude as
\begin{align}
	\rfun{\mscrA_{\text{hfg}}}{\nu} =
		\sqrt{\rfun{\mathrm{GD}_1}{0, \sigma^2 = \frac{1}{2}; \nu}}\,\sgn \nu =
		\pp^{-1/4} \ee^{-\frac{\nu^2}{2}}\sgn \nu\,.
	\label{eq:amp-hfg-10}
\end{align}
%see fig.\ \ref{fig:linear-prob-lin,fig:linear-prob-log}.
The corresponding wave-packet, which is constructed numerically, is plotted in 
fig.\ \ref{fig:linear-wp-10-gau}. One sees that it indeed resembles that with 
$\mscrA_{\text{lin}}$ in fig.\ \ref{fig:linear-wp-10-lin}.

One may ask about a possible classical correspondence of this 
wave-packet, which many other wave-packets do have.
Generally speaking, the 
familiar scenario would be that the wave-packet is constructed by superposing 
mode functions with quantum number $\nu \in \BbbR$ by a normal Gaussian 
amplitude, that is centred at $\nu_0$. Then the claim is that, this wave-packet 
corresponds to the classical trajectory with a classical first-integral 
$\propto \nu$, see also \cite{Andrianov2018wdx,Kiefer2019}. This approach is
not viable here, since the amplitude is by no means a normal Gaussian one.
We will focus on the issue of digging a classical trajectory out of a 
generic wave-packet in section \ref{sec:ridge-line},
but before that, let us revisit the traditional WKB approach to derive the classical trajectory.

%123456789012345678901234567890123456789
\subsection{Narrow WKB Gaussian wave-packet}
\label{ssec:packet-gaussian}
%1234567890123456789012345678901234567890123456789012345678901234567890123456789

In this section \ref{ssec:packet-gaussian}, we 
study a special case, 
in which the wave-packet is constructed by superposing the WKB mode functions 
with a narrow Gaussian amplitude. The mathematical result confirms the 
heuristic idea, that such a wave-packet peaks near the classical trajectory, 
which shares the same integral constant as the centre of the Gaussian amplitude.

\begin{figure}
	\begin{center}
	\begin{subfigure}[t]{.49\textwidth}
		\includegraphics{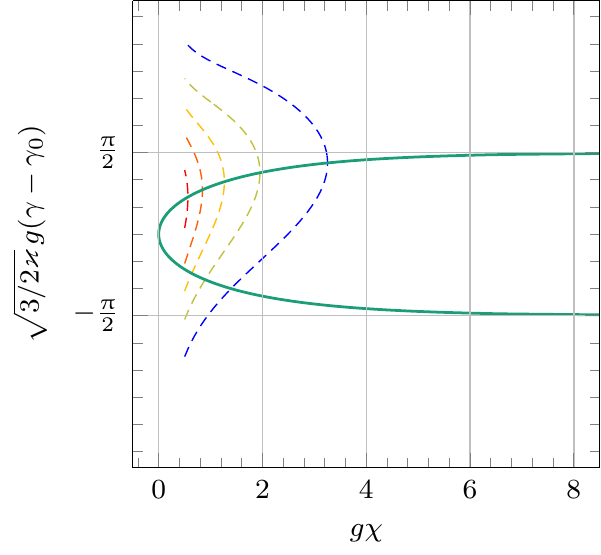}
		\caption{$\rbr{-, +}$ with $S_+$}
		\label{fig:wkb-wp-10-jp}
	\end{subfigure}
	\begin{subfigure}[t]{.49\textwidth}
		\includegraphics{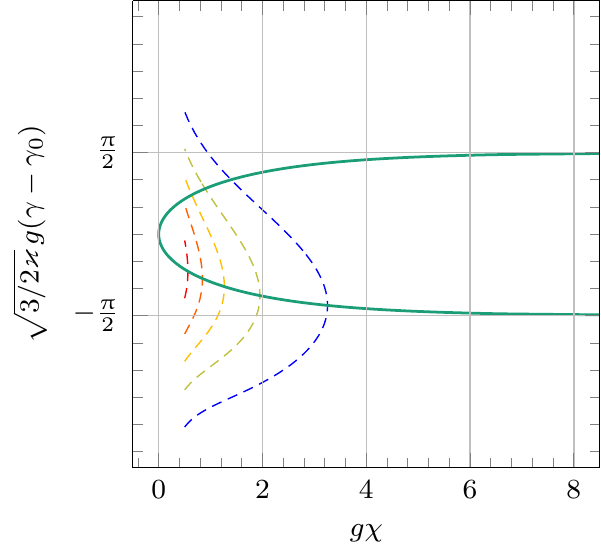}
		\caption{$\rbr{-, +}$ with $S_-$}
		\label{fig:wkb-wp-10-jm}
	\end{subfigure}\\
	\begin{subfigure}[t]{.49\textwidth}
		\includegraphics{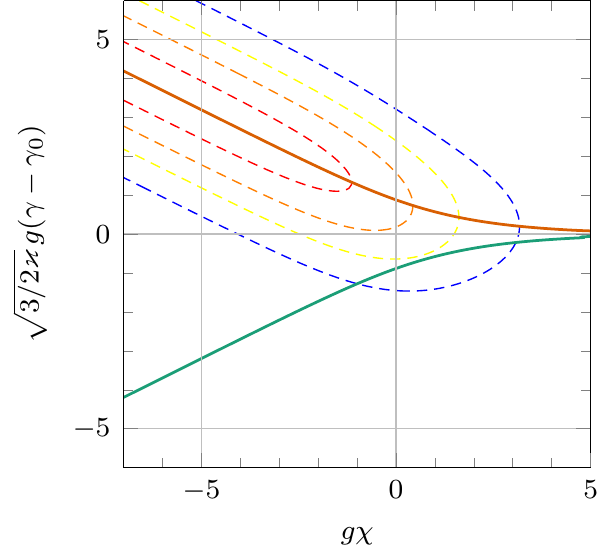}
		\caption{$\rbr{+, -}$ with $S_+$}
		\label{fig:wkb-wp-10-fp}
	\end{subfigure}
	\begin{subfigure}[t]{.49\textwidth}
		\includegraphics{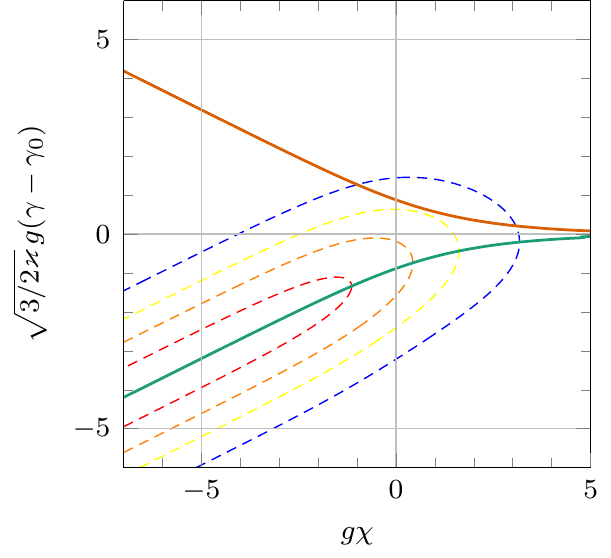}
		\caption{$\rbr{+, -}$ with $S_-$}
		\label{fig:wkb-wp-10-fm}
	\end{subfigure}\\
	\begin{subfigure}[t]{.49\textwidth}
		\includegraphics{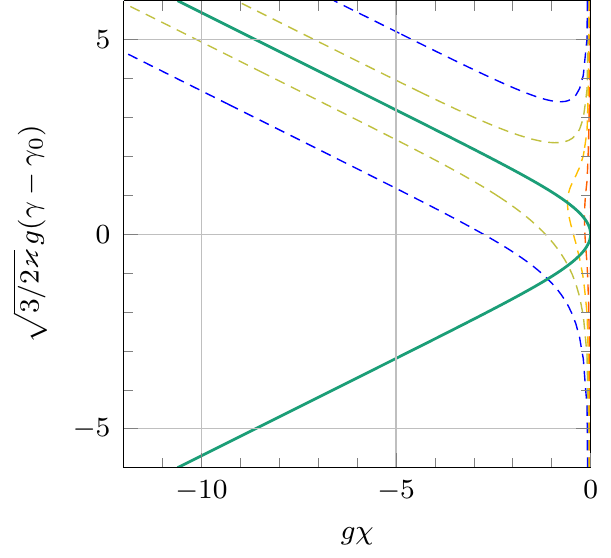}
		\caption{$\rbr{+, +}$ with $S_+$}
		\label{fig:wkb-wp-10-kp}
	\end{subfigure}
	\begin{subfigure}[t]{.49\textwidth}
		\includegraphics{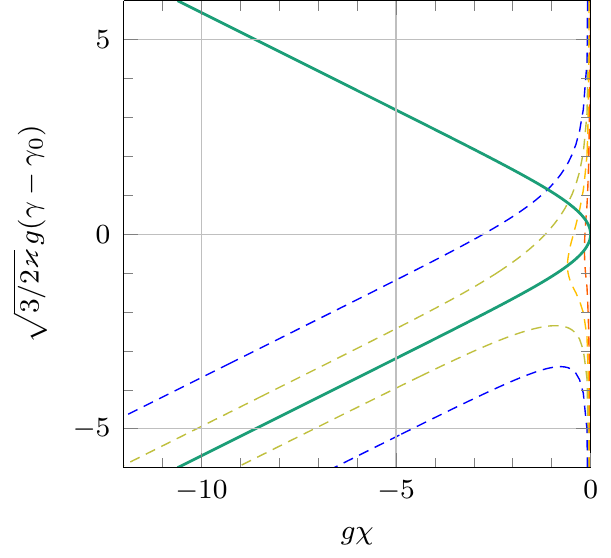}
		\caption{$\rbr{+, +}$ with $S_-$}
		\label{fig:wkb-wp-10-km}
	\end{subfigure}
	\end{center}
	\caption[Narrow Gaussian wave-packets of the WKB mode functions]%
	{Narrow Gaussian wave-packets of the WKB mode functions with $S_{\pm}$ by
	eq.\ \eqref{eq:packet-gaussian-87} as dashed contours, the expressions of which 
	are listed in table \ref{tab:wkb-wp-10}. One sees that for each $S_\pm$, the
	wave-packet peaks around one asymptotic branch of the classical trajectory, 
	which fails to hold near the turning point. Moreover, for the $\rbr{+, -}$- 
	and $\rbr{+, +}$-cases, where $g\chi \to -\infty$
	is a region that the corresponding Bessel functions are sinusoidal,
	the wave-packets form uniform tubes near the classical trajectory.
	For the $\rbr{-, +}$- and $\rbr{+, -}$-cases, where $g\chi \to +\infty$
	is a region that the corresponding Bessel functions decay exponentially
	in amplitude, the wave-packets also decay.\label{fig:wkb-wp-10}}
	%See also fig.\ \ref{fig:wav-fun-10}.
\end{figure}

We begin with the two-dimensional case eq.\ \eqref{eq:prototype-11}, so that the WKB wave function reads (see appendex \ref{app:prototype-wkb})
\begin{align}
	\rfun{\psi}{q^1, q^2; \alpha} \approx \sqrt{D}\,
	\sfun{\exp}{\frac{\ii}{\hslash}
	\rbr{\rfun{S}{q^1, q^2; \alpha} - \alpha \beta}}\,,
\end{align}
where the additional phase $\alpha\beta$ will become clear soon.
The Gaussian wave-packet is the result of
\begin{subequations}
\begin{align}
	\rfun{\varPsi}{q^1, q^2; \alpha, \sigma} &=
	\int \dd \Alpha\,
	\rfun{\psi}{q^1, q^2; \Alpha}
	\rfun{\mathrm{GD}_1}{\alpha, \sigma^2; \Alpha}^{1/2}\,,
	\label{eq:packet-gaussian-50a} \\
	\rfun{\mathrm{GD}_1}{\alpha, \sigma^2; \Alpha} &\coloneqq
	\frac{\rfun{\exp}{-\frac{1}{2} \sigma^{-2}
		\rbr{\Alpha-\alpha}^2}}{\sqrt{2\pp \sigma^2}} \,.
	\label{eq:packet-gaussian-50b}
\end{align}
\end{subequations}
Applying Taylor's theorem to the exponent of the integrand in 
eq.\ \eqref{eq:packet-gaussian-50a} with respect to $\Alpha$ at $\alpha$ gives
\begin{align}
	&\quad\, \rfun{\psi}{q^1, q^2; \Alpha}
		\rfun{\mathrm{GD}_1}{\alpha, \sigma; \Alpha}^{1/2}
	\nonumber \\
	& = \sfun{\exp}{\ii d^{(0)}_1 +
		\ii \rbr{\Alpha - \alpha} d^{(1)}_1 - \frac{1}{2}
		\rbr{\Alpha - \alpha}^2 d^{(2)}_1}
		\rfun{g}{\Alpha}\,,
		\label{eq:packet-gaussian-75a}
\end{align}
where
\begin{subequations}
\begin{align}
	d^{(0)}_1 &\coloneqq \frac{1}{\hslash}\rbr{\rfun{S}{q^1, q^2; \alpha}
		- \alpha \beta}\,,
	\\
	d^{(1)}_1 &\coloneqq
		\frac{1}{\hslash}\rbr{\partial_{\alpha} S - \beta}\,,
	\\
	d^{(2)}_1 &\coloneqq
		\frac{1}{2} \sigma^{-2}
		- \frac{\ii}{\hslash}\partial_{\alpha}^{2} S\,;
	\\
	\rfun{g}{\Alpha} &\coloneqq
		\sqrt{D}\,\rfun{\exp}{\rfun{h}{\Alpha}\rbr{\Alpha - \alpha}^2}\,,
	\qquad
	\rfun{h}{\alpha} = 0\,.
\end{align}
\end{subequations}
If $d^{(2)}_1$ dominates in eq.\ \eqref{eq:packet-gaussian-75a}, i.e.\ 
$\vbr{d^{(2)}_1} \gg 1$, the integral in eq.\ \eqref{eq:packet-gaussian-50a} can be
estimated by the \emph{stationary phase method} 
\cite{dabrowski2006,Andrianov2018wdx}. 
This can be realised if $\sigma^{-2} \gg \hslash^{-1} \partial_\alpha^2 S$,
which means that the wave-packet is constructed to be \emph{narrow}. 
The result is
\begin{align}
	\rfun{\varPsi}{q^1, q^2; \alpha, \sigma} \approx \rbr{2\pp}^{1/4}
		\rbr{\frac{D}{\sigma d^{(2)}_1}}^{1/2}
	\sfun{\exp}{\ii d^{(0)}_1 - \frac{\rbr{d^{(1)}_1}^{2}}{2d^{(2)}_1}}\,,
\end{align}
and the corresponding Schr\"odinger density reads
\begin{align}
	\rho = \rfun{\rho}{q^1, q^2; \alpha, \sigma} = \vbr{\varPsi}^2 =
	\sqrt{2\pp}\,\frac{D}{\sigma \vbr{d^{(2)}_1}}\,
	\sfun{\exp}{-\frac{\Re d^{(2)}_1}%
		{\vbr{d^{(2)}_1}^2}\rbr{d^{(1)}_1}^{2}}\,.
	\label{eq:packet-gaussian-87}
\end{align}
Given that $D$, $d^{(2)}_1$ and $\rfun{\Re}{d^{(2)}_1}$ vary slowly with 
respect to $\rbr{q^1, q^2}$, the peak of $\rho$ dominates near $d^{(1)}_1 = 0$, 
i.e.\ $\partial_\alpha S = \beta$ (c.f.\ eq.\ \eqref{eq:prin-cons-int-20}), which is 
just the classical trajectory. Narrow Gaussian wave-packets of $\rbr{-, +}$, 
$\rbr{+, -}$ and $\rbr{+, +}$ cases are summarized in table \ref{tab:wkb-wp-10} and 
plotted in fig.\ \ref{fig:wkb-wp-10}.

\begin{table}
\begin{center}
\begin{tabular}{l|r@{$\cdot$}l}
\toprule
$\rbr{\mitsansl, \mitsanss \mitsansv}$ & 
$\sqrt{2\pp} \frac{D}{\sigma\vbr{d_1^{(2)}}}$ &
$\sfun{\exp}{-\frac{\Re d_1^{(2)}}{\vbr{d_1^{(2)}}} \rbr{d_1^{(1)}}^2 }$
\\
\midrule
$\rbr{-, -}$ & \multicolumn{2}{c}{no solution} \\
$\rbr{-, +}$ &
$\frac{2 \sqrt{2 \pp }\, \sigma}{\sqrt{+x^2-\nu ^2+4 \sigma ^4}}$ &
$\sfun{\exp}{-\frac{\sigma ^2 \rbr{+x^2-\nu ^2}}{x^2-\nu ^2+4 \sigma ^4}
		\rbr{\sqrt{\frac{3}{2\varkappa}}\,g \rbr{\gamma-\gamma_0} \mp
			\arccos\frac{\nu }{x}}^2}$\\
$\rbr{+, -}$ &
$\frac{2 \sqrt{2 \pp }\, \sigma}{\sqrt{+x^2+\nu ^2+4 \sigma ^4}}$ &
$\sfun{\exp}{-\frac{\sigma ^2 \rbr{+x^2+\nu ^2}}{x^2+\nu ^2+4 \sigma ^4}
		\rbr{\sqrt{\frac{3}{2\varkappa}}\,g \rbr{\gamma-\gamma_0} \mp
			\arcsinh\frac{\nu }{x}}^2}$\\
$\rbr{+, +}$ &
$\frac{2 \sqrt{2 \pp }\, \sigma}{\sqrt{-x^2+\nu ^2+4 \sigma ^4}}$ &
$\sfun{\exp}{-\frac{\sigma ^2 \rbr{-x^2+\nu ^2}}{x^2+\nu ^2+4 \sigma ^4}
		\rbr{\sqrt{\frac{3}{2\varkappa}}\,g \rbr{\gamma-\gamma_0} \mp
			\arccosh\frac{\nu }{x}}^2}$\\
\bottomrule
\end{tabular}
\end{center}
\caption[Narrow Gaussian wave-packets of the WKB mode functions]{%
	Narrow Gaussian wave-packet of the WKB mode functions with $S_{\pm}$ by
		eq.\ \eqref{eq:packet-gaussian-87}, which are plotted in 
		fig.\ \ref{fig:wkb-wp-10}. }
\label{tab:wkb-wp-10}
\end{table}

The above result in two dimensions can easily be generalised to higher 
dimensions. Consider the WKB mode function
\begin{align}
	\rfun{\psi}{q^i; \alpha_k} \approx \sqrt{D}\,
		\sfun{\exp}{\frac{\ii}{\hslash}
			\rbr{\rfun{S}{q^1\ldots q^n; \alpha_1\ldots \alpha_m}
				- \sum_{k=1}^{m}\alpha_k \beta_k}}\,,
\end{align}
where $m = n-1$ is the number of integral constants.

Choosing a non-degenerate $m$-dimensional Gaussian amplitude leads
to the Gaussian wave-packet
\begin{subequations}
\begin{align}
	\rfun{\varPsi}{q^i; \alpha_j, \varSigma_{jk}} &=
	\int \dd \Alpha_1\ldots \dd \Alpha_{m}\,
	\rfun{\psi}{q^i; \Alpha_k}
	\rfun{\mathrm{GD}_m}{\alpha_k, \varSigma_{kl}; \Alpha_k}^{1/2}\,,
	\label{eq:eq:packet-gaussian-100a}
\end{align}
where
\begin{align}
	\rfun{\mathrm{GD}_m}{\alpha_k, \varSigma_{kl}; \Alpha_k} &\coloneqq
		\frac{\sfun{\exp}{-\frac{1}{2} \sum_{k,l=1}^{m}
			\rbr{\varSigma^{-1}}_{kl}
				\rbr{\Alpha-\alpha}_k \rbr{\Alpha-\alpha}_l}}%
			{\sqrt{\rbr{2\pp}^{m} \det \varSigma}}
\label{eq:packet-gaussian-100}
\end{align}
\end{subequations}
is the probability density function of the multivariate Gaussian distribution
\cite[ch.\ 5]{Gut2009},
$m = n-1$, and $\varSigma$ is the non-degenerate, positive definite covariance 
matrix. The integral in eq.\ \eqref{eq:eq:packet-gaussian-100a} can also be estimated
by the stationary phase method as
\begin{align}
	\begin{split}
		\rfun{\varPsi}{q^i; \alpha_k, \varSigma_{kl}}
		&\approx
			\rbr{\frac{\rbr{2\pp}^{m}}{\det \varSigma}}^{1/4}
				\rbr{\frac{D}{\det d^{(2)}_m}}^{1/2}
		\\
		&\quad\,\cdot
		\rfun{\exp}{\ii d^{(0)}_m - \frac{1}{2} \sum_{k,l}
			\rbr{d^{(2)}_m}_{kl} \rbr{d^{(1)}_m}_{k} \rbr{d^{(1)}_m}_{l}}\,,
	\end{split}
\end{align}
where
\begin{subequations}
\begin{align}
	d^{(0)}_m &\coloneqq \frac{1}{\hslash}
		\rbr{\rfun{S}{q^i; \alpha^k} - \sum_{k=1}^{m}\alpha_k \beta_k}\,,
	\\
	\rbr{d^{(1)}_m}_{k} &\coloneqq
		\frac{1}{\hslash}\rbr{\partial_{\alpha_k} S - \beta_k}\,,
	\\
	\rbr{d^{(2)}_m}_{kl} &\coloneqq
		\rbr{\frac{1}{2} \varSigma^{-1}
			- \frac{\ii}{\hslash}\mathrm{Hess}_{\alpha}\, S}_{kl}\,;
	\\
	\rbr{\mathrm{Hess}_{\alpha}\, S}_{kl} &\coloneqq
		\partial_{\alpha_k}\,\partial_{\alpha_l}S\,.
\end{align}
\end{subequations}
The Schr\"odinger density of the wave-packet reads
\begin{align}
\begin{split}
	\rho &= \rfun{\rho}{q^i, \alpha_k, \varSigma_{kl}} = \vbr{\varPsi}^2 
	\\
	&=
	\sqrt{\frac{\rbr{2\pp}^{m}}{\det \varSigma}} \frac{D}{\det d^{(2)}_m}\,
	\sfun{\exp}{-\rfun{\Re}{\sum_{k,l} %
		\rbr{d^{(2)}_m}_{kl} \rbr{d^{(1)}_m}_{k} \rbr{d^{(1)}_m}_{l}}}\,.
\label{eq:packet-gaussian-187}
\end{split}
\end{align}
The corresponding classical trajectory is
$\rbr{d^{(1)}_m}_{k} = 0$, or $\beta_k = \partial_{\alpha_k} S$, which is 
identical to eq.\ \eqref{eq:prin-cons-int-10}.

%1234567890123456789012345678901234567890123456789012345678901234567890123456789

Therefore, we can be
confident that a classical universe is likely to emerge from a quantum 
wave-packet constructed by a narrow Gaussian amplitude, and in regions 
where the WKB approximation is good. The amplitudes near the peak also seem
to be constant. Departure from classical theory is expected where these conditions 
are violated, for example when the wave-packet spreads (and becomes wider), 
is damped (and the amplitude becomes smaller), or near the classical turning
point (and the WKB approximation fails).

The idea of the \enquote{peak} of a wave-packet, that was used in 
eqs.\ \eqref{eq:packet-gaussian-87} and \eqref{eq:packet-gaussian-187}, is heuristic. If a 
wave-packet does not have a form as in 
eqs.\ \eqref{eq:packet-gaussian-87} and \eqref{eq:packet-gaussian-187}, the heuristic idea
does not easily apply, which has already happened in eq.\ \eqref{eq:linear-pac-Psi}.
One needs a mathematical description for this idea, which will be studied in
section \ref{sec:ridge-line}. One will see that in the contour approach of ridge-lines,
as well as in the simple first-derivative test, the classical trajectories in
eqs.\ \eqref{eq:packet-gaussian-87} and \eqref{eq:packet-gaussian-187} can be confirmed.

%1234567890123456789012345678901234567890123456789012345678901234567890123456789
\section{Stationary wave-packets}
\label{sec:sta-wp}
%1234567890123456789012345678901234567890123456789012345678901234567890123456789

In quantum cosmology, the usual way of constructing a wave-packet is linearly
superposing the complete integrals $\psi_{\nu}$, containing constants $\nu$, of 
the Wheeler--DeWitt equation
which is comparable to the stationary 
Schr\"odinger equation in quantum mechanics,  
\begin{align}
H \psi_\nu = E \psi_\nu\,,
\label{eq:stat-schroe-10}
\end{align}
the solution $\psi_\nu$ to which is called the wave function of a 
\emph{stationary state}, where $\nu$ is another quantum number that marks
different states in a degenerate level. If one writes $H_\perp = H - E$ and 
fixes the energy level $E$, eq.\ \eqref{eq:stat-schroe-10} becomes 
$H_\perp \psi_\nu = 0$, which looks identical to eq.\ \eqref{eq:MWDW-10}. In this 
resemblance, constructing a wave-packet corresponds to the superposition of
degenerate stationary states in the same energy level, the result of which is 
also an energy eigenstate of the same level.

We will call such a quantum wave-packet a \emph{stationary wave-packet}, that
encompasses both conventional quantum mechanics and the 
Wheeler--DeWitt quantum cosmology. Relating a tentative theory of quantum
gravitation to quantum mechanics can lead to analogue models, which has been 
realised in the study of black holes 
\cite{Unruh1981,Weinfurtner2010,Steinhauer2015,Barcelo2018} and quantum field 
theory in curved space-time \cite{Boettcher2019,Weinfurtner2005}. For a review 
of analogue gravitation, see \cite{Barcelo2005}.

On 
the other hand, we noticed that the Rydberg or highly-excited atom, has indeed 
a description of such a superposition as a wave-packet
\cite{Gallagher1994,Lim2013,Sibalic2018}. Independent of this experimental 
aspect, in section \ref{ssec:qm-2dha} we introduce the two-dimensional hydrogen atom 
as a toy model, and then construct stationary wave-packets in 
section \ref{ssec:qm-swp}. Meanwhile, 
we discuss the
choice of superposition amplitudes, arguing in favour of Gaussian, binomial
and Poisson amplitudes, etc., which maximises the entropy. In the end, we
turn to the study of the classical limit, and verify the correspondence
principles in section \ref{ssec:qm-rl}.

%123456789012345678901234567890123456789
\subsection[Two-dimensional hydrogen atom]%
{\boldmath Two-dimensional hydrogen atom}
\label{ssec:qm-2dha}
%123456789012345678901234567890123456789

Consider a spinless non-relativistic two-dimensional hydrogen atom, described 
by the action
\begin{align}
	S = \int \dd t\, \sbr{
			\frac{m}{2} \rbr{\dot{\varrho}^2 + \varrho^2 \dot{\varphi}^2}
		+ \frac{\alpha}{\varrho}}\,,
	\qquad
	\alpha > 0
	\label{eq:2d-hydrogen-10}
\end{align}
in polar coordinates $\rbr{\varrho, \varphi}$. The classical trajectory can be 
solved in terms of the conserved energy and angular momentum $\rbr{E, L}$ as
\begin{align}
	\varrho = \frac{L^2}%
	{m \alpha + \sqrt{m \rbr{2E L^2 + m\alpha^2}}\,
		\rfun{\cos}{\varphi - \varphi_0}}\,.
	\label{eq:2dhyd-cla-traj-10}
\end{align}
For $E < 0$, the system is bounded, and the trajectory is an ellipse. Fixing
$\varphi_0 = 0$, the trajectory passing through
$\rbr{\varrho, \varphi} = \rbr{\varrho_0, 0}$ and $\rbr{\varrho_\pp, \pp}$ can
be worked out in terms of
\begin{align}
	E = -\frac{\alpha}{\varrho_0 + \varrho_{\pp}} < 0\,,
	\qquad
	L = \pm \sqrt{\frac{2m\alpha}{\varrho_0^{-1} + \varrho_{\pp}^{-1}}}\,.
	\label{eq:2dhyd-cla-traj-20}
\end{align}

Upon canonical quantisation, the stationary Schr\"odinger equation reads
\begin{align}
	\rbr{-\frac{\hslash^2}{2m}\nabla^2 - \frac{\alpha}{\varrho}} 
	\rfun{\psi}{\varrho, \varphi} = E \rfun{\psi}{\varrho, \varphi}\,,
	\label{eq:stat-schroe-20}
\end{align}
where the Laplace--Beltrami operator
\begin{align}
	\nabla^2 \coloneqq \partial_{\varrho}^2 + \varrho^{-1}\partial_{\varrho}
	- \frac{1}{\hslash^2\varrho^{2}} L^2\,, \qquad
	L \coloneqq -\ii \hslash\,\partial_{\varphi}
	\label{eq:lap-bel-10}
\end{align}
is chosen. %In section \ref{ssec:qm-2dha-quant} we show that 
The stationary wave 
functions, with definite main and angular quantum numbers, are
\begin{subequations}
\begin{align}
	\rfun{\psi_{nl}}{\xi, \varphi} &=
	\rfun{\Rho_{nl}}{\xi}\rfun{\varPhi_l}{\varphi}\,,
	\label{eq:2dhyd-eneein-20} \\
	\rfun{\Rho_{nl}}{\xi} &=
	N_{nl} \xi^{\vbr{l}} \ee^{-\xi/2} \rfun{G_{nl}}{\xi}\,,
	\\
	N_{nl} &= \frac{1}{\rbr{2\vbr{l}}!}\rbr{\frac{\rbr{n+\vbr{l}}!}{\rbr{2n+1}%
	\rbr{n-\vbr{l}}!}}^{1/2}\,,
	\label{eq:normal-2d-10} \\
	\rfun{\varPhi_l}{\varphi} &= \rbr{2\pp}^{-1/2} \ee^{\ii l \varphi},
	\qquad
	l = 0, \pm 1, \pm 2, \ldots\,,
	\label{eq:2dhyd-eneein-10}	
\end{align}
where
\begin{align}
	\xi \coloneqq \beta_n \varrho\,,\qquad
	\beta_n \coloneqq \frac{2m\alpha}{\hslash^2}\rbr{n+\frac{1}{2}}^{-1}
	\label{eq:2d-xi-10}
\end{align}
\end{subequations}
are the \emph{dimensionless radial coordinate}, and $G_{nl}$
can be given in terms of 
a Kummer's \cite{Kummer1837a} confluent hypergeometric function
\cite[sec.\ 13.2]{NIST:DLMF}, Sonin's \cite[sec.\ 40]{Sonine1880}
associated Laguerre polynomial \cite[eq.\ (18.11.2)]{NIST:DLMF}, or a 
Whittaker function \cite[eq.\ (13.14.4)]{NIST:DLMF} as
\begin{subequations}
\begin{alignat}{3}
	\rfun{G}{\xi} &=
	\rfun{{}_1 F_{1}}{\vbr{l}-n, 2\vbr{l}+1, \xi}\,	N_{nl}
	\label{eq:2d-gnl-10} \\
	&= \rfun{L^{\rbr{a}}_{\mu}}{\xi}
	\frac{a!}{\rbr{\mu+1}_{2\vbr{l}}}\,
	N_{nl}\qquad
	&a &= 2\vbr{l}\,, \quad
	\mu = n-\vbr{l}\,;
	\label{eq:2d-gnl-20} \\
	&= \rfun{M_{\nu, \vbr{l}}}{\xi} \xi^{-\vbr{l}}\ee^{\xi/2}\,
	N_{nl}\qquad
	&\nu &= n+\frac{1}{2}\,,
	\label{eq:2d-gnl-30}
\end{alignat}
\end{subequations}
where $\rbr{a}_n \coloneqq a\rbr{a-1}\ldots\rbr{a-n+1}$ is the Pochhammer's
\cite{Pochhammer1888} symbol \cite[sec.\ 5.2(iii)]{NIST:DLMF}.
Note that 
eq.\ \eqref{eq:normal-2d-10} is chosen such that eq.\ \eqref{eq:2dhyd-eneein-20} is 
normalised with respect to $\xi$, rather than $\varrho$. The energy levels for 
the bounded states are
\begin{align}
	E_n \coloneqq -\frac{m\alpha^2}{2\hslash^2} \rbr{n+\frac{1}{2}}^{-2}\,.
	\label{eq:2d-ene-10}
\end{align}

The normalisation condition for scattering states $E \ge 0$ does not lead to a 
closed-form expression for the normalisation factor, see e.g.\ 
\cite[eq.\ (2.28)]{Yang_1991}. For simplicity, we focus on the case $E < 0$
in the following.

%123456789012345678901234567890123456789
\subsection{\boldmath Stationary wave-packets for the hydrogen atom}
\label{ssec:qm-swp}
%123456789012345678901234567890123456789

\begin{figure}
	\centering
	\begin{subfigure}{.49\linewidth}
		\centering
		\includegraphics{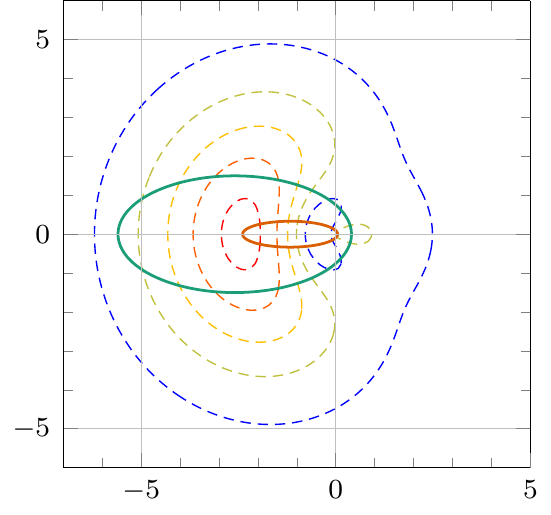}
		\caption{$n=1$, $q=\frac{3}{4}n$
			\label{fig:2dhyd-sta-pac-10-10}}
	\end{subfigure}
	\begin{subfigure}{.49\linewidth}
		\centering
		\includegraphics{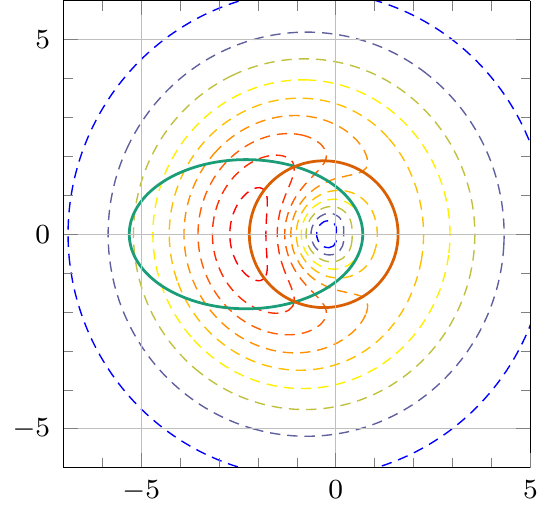}
		\caption{$n=1$, $q=\frac{23}{24}n$
			\label{fig:2dhyd-sta-pac-10-20}}
	\end{subfigure}\\[10pt]
	\begin{subfigure}{.49\linewidth}
		\centering
		\includegraphics{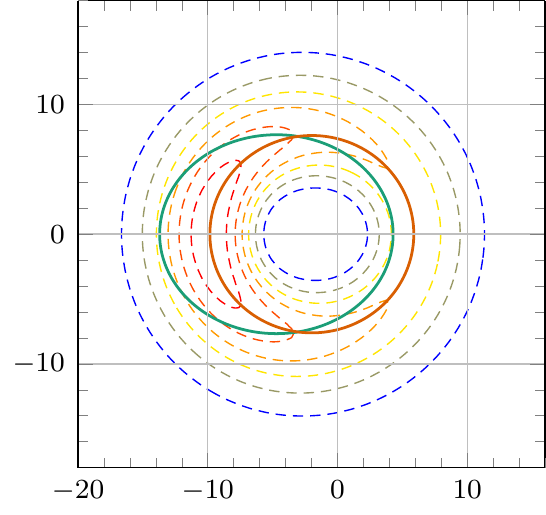}
		\caption{$n=4$, $q=\frac{23}{24}n$
			\label{fig:2dhyd-sta-pac-10-30}}
	\end{subfigure}
	\begin{subfigure}{.49\linewidth}
		\centering
		\includegraphics{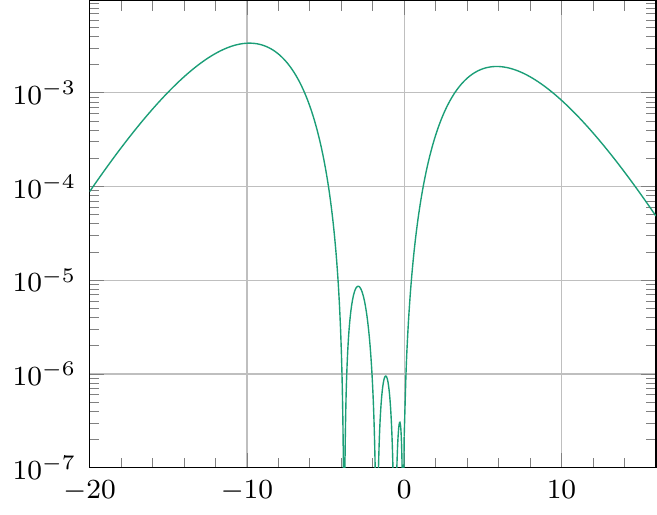}
		\caption{$n=4$, $q=\frac{23}{24}n$, $\sin\varphi = 0$
			\label{fig:2dhyd-sta-pac-10-30-x}}
	\end{subfigure}
\caption[Stationary wave-packets of the two-dimensional hydrogen atom 1]%
{Stationary wave-packets $\vbr{\rfun{\varPsi_{nq}}{\xi, \varphi}}^2$ of the 
two-dimensional hydrogen atom. The green lines denote a classical trajectory 
in eq.\ \eqref{eq:2dhyd-cla-traj-10} with $\varphi_0 = 0$, $E = E_n$ and 
$L = q\hslash$. The orange lines are the classical trajectories passing through
the two highest peaks of the wave-packet, with the integral constants 
$\rbr{E, L}$ given by eq.\ \eqref{eq:2dhyd-cla-traj-20}. Apparently, the
green line in fig.\ \ref{fig:2dhyd-sta-pac-10-20} fits the orange line better
than in fig.\ \ref{fig:2dhyd-sta-pac-10-10}, but worse than in
fig.\ \ref{fig:2dhyd-sta-pac-10-30}. In fig.\ \ref{fig:2dhyd-sta-pac-10-30-x} we show
the normal projection of fig.\ \ref{fig:2dhyd-sta-pac-10-30} on the $\sin\varphi=0$ 
line (in logarithm scale). One sees that there are multiple maxima; the highest
two were chosen for plotting fig.\ \ref{fig:2dhyd-sta-pac-10-30}.
\label{fig:2dhyd-sta-pac-10}}
\end{figure}

For bounded states of the two-dimensional hydrogen atom in 
eq.\ \eqref{eq:2d-hydrogen-10}, one fixes $E$ or $n$ and chooses a probability 
amplitude for different $l$'s to construct a stationary wave-packet,
\begin{align}
	\varPsi_{nq} \coloneqq \sum_{k=-n}^{n} A_{nk;q} \psi_{nk}\,.
	\label{eq:wave-packet-10}
\end{align}
We would like to find a choice for the $A_{nk;q}$'s, such that the expectation 
value of angular momentum
\begin{align}
\rbr{\varPsi_{nq}, L\,\varPsi_{nq}} = q\hslash\,,
\end{align}
where $q \in \sbr{-n, n}$, $q \in \mathbb{R}$.
Since $k \in \sbr{-n, n} \cap \mathbb{Z}$, a \enquote{natural} choice for the
probability masses seems to be the \emph{binomial distribution},
where the probability mass function is
\begin{subequations}
\begin{align}
	\rfun{\mathrm{BD}}{k; u, s} &\coloneqq
	\binom{u}{k} s^k \rbr{1-s}^{u-k}\,,
	\label{eq:qm-swp-bd-10} \\
	\binom{u}{k} &\coloneqq \frac{u!}{k!\rbr{u-k}!}\,,
	\\
	k &=0, 1, \ldots, u,
	\qquad
	s \in \sbr{0,1}\,.
\end{align}
\end{subequations}
In our case, the amplitude satisfies
\begin{subequations}
\begin{align}
	\vbr{A_{nk;q}}^2 &= \rfun{\mathrm{BD}}{n+k, 2n, \frac{n+q}{2n}}
	\nonumber \\
	&= \rbr{2n}^{-2n} \rbr{n-q}^{n-k}\rbr{n+q}^{n+k} \binom{2n}{n+k}\,.
\end{align}
The most naive choice
\begin{align}
	A_{nk;q} = \sqrt{\rfun{\mathrm{BD}}{n+k, 2n, \frac{n+q}{2n}}}
	\label{eq:binomial-20}
\end{align}
\end{subequations}
leads to stationary wave-packets that \enquote{peak around} a classical 
trajectory for $\vbr{q} \lesssim n$, see fig.\ \ref{fig:2dhyd-sta-pac-10}.

%123456789012345678901234567890123456789
\subsection{Ridge-line of a wave-packet and the correspondence principles}
\label{ssec:qm-rl}
%1234567890123456789012345678901234567890123456789012345678901234567890123456789

In quantum cosmology, people argue that the ridge-line of a wave-packet peaks
along a classical trajectory \cite{Halliwell2009}. This would be more 
convincing if the statement also holds for the stationary wave-packets in
conventional quantum mechanics.

For the binomial wave-packets here, defined by 
eqs.\ \eqref{eq:2dhyd-eneein-20}, \eqref{eq:2dhyd-eneein-10},
\eqref{eq:wave-packet-10} and \eqref{eq:binomial-20},
we approximate the ridge by finding the two highest peaks of the wave-packet,
and find the elliptic classical trajectory passing them, see 
fig.\ \ref{fig:2dhyd-sta-pac-10}. The approximate ridge-line
is described by the integral constants $\rbr{E_{\text{ar}}, L_{\text{ar}}}$
given by eq.\ \eqref{eq:2dhyd-cla-traj-20}.

\begin{figure}
	\centering
	\begin{subfigure}{.49\linewidth}
		\centering
		\includegraphics{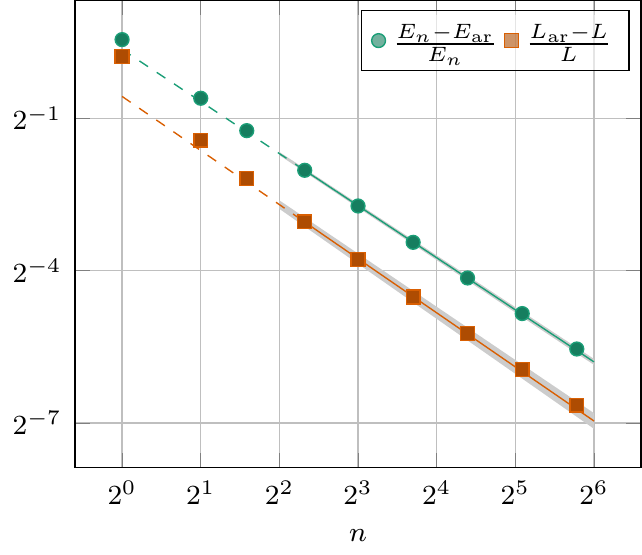}
		\subcaption{Bohr's, $\frac{l}{n} \equiv \frac{23}{24}$.
			\label{fig:2dhyd-sta-pac-20-bohr}}
	\end{subfigure}
	\begin{subfigure}{.49\linewidth}
		\centering
		\includegraphics{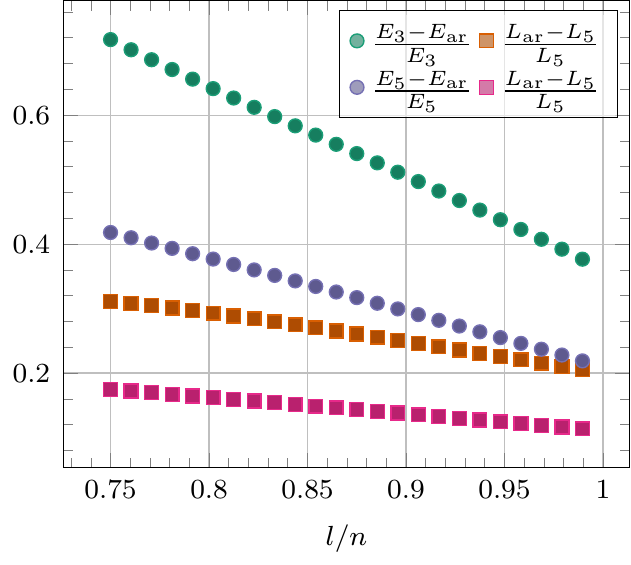}
		\subcaption{Ours, $n = 3$ and $5$.
			\label{fig:2dhyd-sta-pac-20-our}}
	\end{subfigure}
	\caption[Correspondence principles]{Correspondence principles shown in 
terms of the difference between $\rbr{E_\text{ar}, L_\text{ar}}$ and 
$\rbr{E_n, q\hslash}$, where the former with subscript $_{\text{ar}}$ 
denotes the integral constants that give a trajectory passing through the 
two highest peaks of the binomial wave-packet (fig.\ \ref{fig:2dhyd-sta-pac-10}). 
In fig.\ \ref{fig:2dhyd-sta-pac-20-bohr}, the difference vanishes as 
$n \to +\infty$, which is accordance with Bohr. The solid line is the best fit 
with the generalised linear model \cite{Nelder1972} 
$y = \rfun{g^{-1}}{\beta_0 + \beta_1 \ln n}$ with $\rfun{g}{y} = \ln y$. In 
fig.\ \ref{fig:2dhyd-sta-pac-20-our}, the difference becomes smaller as $q \to n^-$, 
but will not vanish; this correspondence phenomenon is relevant in quantum 
cosmology.
\label{fig:2dhyd-sta-pac-20}}
\end{figure}

One sees that this approximation is good as $n$ increases, which fits Bohr's
correspondence principle \cite{Bohr1920}, stating that the quantum system 
reproduces its classical behaviour in the limit of large \emph{main} quantum 
number $n$. This can be seen in fig.\ \ref{fig:2dhyd-sta-pac-20-bohr}, where one 
fixes $q/n$ and observes the relative difference between 
$\rbr{E_\text{ar}, L_\text{ar}}$ and $\rbr{E_n, L}$ vanishes polynomially as 
$n \to +\infty$.

In our application, on the other hand, we are more concerned with fixed $n$ or 
$E_n$, and varying $q$. In this case, the ridge-line gets closer to the 
classical trajectory as the effective \emph{angular} quantum number 
$q \to n^-$, in the sense that the relative differences between 
$\rbr{E_\text{ar}, L_\text{ar}}$ and $\rbr{E_n, q\hslash}$ become smaller in 
the aforementioned limit, see fig.\ \ref{fig:2dhyd-sta-pac-20-our}. The differences, 
however, will not vanish. This correspondence phenomenon is relevant in quantum
cosmology, where the \enquote{main quantum number} is to be fixed, and only the 
other quantum numbers in the degenerate \enquote{energy eigenspace} can change.

%1234567890123456789012345678901234567890123456789012345678901234567890123456789
\section{Ridge-lines of wave-packets}
\label{sec:ridge-line}
%1234567890123456789012345678901234567890123456789012345678901234567890123456789

%1234567890123456789012345678901234567890123456789012345678901234567890123456789
\subsection{The conception of ridge-lines}
\label{ssec:concep-ridge-line}
%1234567890123456789012345678901234567890123456789012345678901234567890123456789

In the remaining 
sections of this paper,
we try to quantify the qualitative arguments in 
the literature, that a classical trajectory can be read off from wave-packets 
in specific forms. Intuitively, one may imagine the profile of a wave-packet
as a terrain in its configuration space, where the hills and valleys are the 
most and least probable places to \enquote{find} the system. In physical 
geography, chains of mountains or hills stretch a distance, where the 
\enquote{highest points} form the \emph{ridge-lines}; conversely, one can 
define the valleys or the \emph{dale-lines} by the \enquote{lowest points}.

The ridge- and dale-lines are in some sense the generalisation of local maxima 
and minima, which are isolated points. The latter are also easier to be solved 
in terms of local extrema as $\nabla \rho = 0$ as necessary but not sufficient 
conditions, and distinguishing them is more involved. One may give a sufficient 
condition when the Hessian is non-singular, but when it is, more works need to 
be done. For simplicity and clearance, we will study the ridge- and dale-lines 
on the equal foot.

The ridge- and dale-lines have been studied by the computer scientists
working on imaging and vision \cite{Koenderink1993,Koenderink1994,Eberly1996}, 
where the ridge- and dale-lines have rich applications, especially in 
two-dimensional \emph{Euclidean geometry}. In physical configuration spaces 
having a higher-dimensional \emph{(pseudo-)Riemannian geometry}, the ridge- and 
dale-lines have not been much used, to our knowledge. In addition, the 
Euclidean experience from computer science also needs to be thought twice.

From now on, we will not use the analogy with terrain any further, which we 
argue as follows. For terrain, the altitude has the dimension of 
length, which is comparable to the dimension of the geographic coordinates. 
For a wave-packet, in contrast, the dimension of its profile is \emph{not} 
comparable to the dimensions of the configuration space coordinates; the 
former might be the inverse of the configuration volume if one has the 
Schr\"odinger normalisation condition in mind,
\begin{align}
\int \dd \text{Vol}\, \vbr{\varPsi}^2 = 1\,,
\end{align}
which is \emph{dependent} on the configuration space coordinates. Based on
these considerations, we shall find an \emph{intrinsic} description of the
ridge-lines of a wave-packet, where the wave-packet is \emph{not} to be 
plotted in an additional dimension.

%123456789012345678901234567890123456789
\subsection{First-derivative test and Hessian matrix}
\label{sec:hr-first}
%1234567890123456789012345678901234567890123456789012345678901234567890123456789

Heuristically, one can simply use the first partial derivative to find the ridge- and dale-lines.
In two dimensions with Cartesian coordinates $\rbr{x, y}$, it reads 
\begin{align}
\rho_{,x} = 0 \qquad \text{or} \qquad \rho_{,y} = 0\,,
\label{eq:hr-first-05}
\end{align}
which is weaker than the extremum condition $\rho_{,x} = 0$ \emph{and} 
$\rho_{,y} = 0$.

Geometrically, eq.\ \eqref{eq:hr-first-05} can be interpreted as a directional 
extremum test, namely to find the extremum with respect to only the $x$- or 
$y$-direction.

\begin{figure}
\begin{center}
\begin{subfigure}{.49\linewidth}
	\begin{center}
\includegraphics{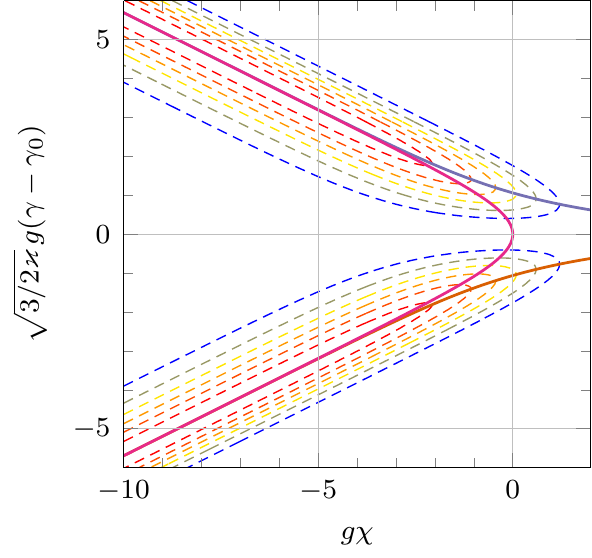}
	\end{center}
	\end{subfigure}
\caption%
	{The first-derivative approach shown with the \enquote{linear} wave-packet in
	eq.\ \eqref{eq:linear-pac-Psi}. 
	The orange and 
	purple lines are the results from eq.\ \eqref{eq:hr-first-20}, whereas the pink 
	line is from eq.\ \eqref{eq:hr-first-10}.}
\label{fig:linear-wp-rl-1der}
\end{center}
\end{figure}

Take the \enquote{linear} wave-packet in eq.\ \eqref{eq:linear-pac-Psi} as an 
example. With $\rho_{\text{lin}} = \vbr{\varPsi_\text{lin}}^2$,
the condition $\partial_\chi \rho_{\text{lin}} = 0$ gives
\begin{align}
\ee^{g \chi}
		\sfun{\cosh}{\sqrt{\tfrac{3}{2\varkappa}}\,g\rbr{\gamma-\gamma_0}}^2
	= \frac{g^2\hslash^2}{8 \mathrm{Vol}_3^2 \vbr{V}}\,.
	\label{eq:hr-first-10}
\end{align}
Compared with eq.\ \eqref{eq:cla-traj-10} and table \ref{tab:trig-10}, eq.\ \eqref{eq:hr-first-10} has
exactly the form of a classical trajectory, with
\begin{align}
p_\gamma^2 = \frac{3 g^2 \hslash^2}{2 \varkappa}\,.
\label{eq:hr-first-15}
\end{align}
On the other hand, the condition $\partial_\gamma \rho_{\text{lin}} = 0$
gives
\begin{align}
\ee^{g \chi}
		\sfun{\cosh}{\sqrt{\tfrac{3}{2\varkappa}}\,g\rbr{\gamma-\gamma_0}}^2
	= \frac{g^2\hslash^2}{8 \mathrm{Vol}_3^2 \vbr{V}}
		\sfun{\coth}{\sqrt{\tfrac{3}{2\varkappa}}\,g\rbr{\gamma-\gamma_0}}^4\,.
	\label{eq:hr-first-20}
\end{align}
Since $\sfun{\coth}{\sqrt{\tfrac{3}{2\varkappa}}\,g\rbr{\gamma-\gamma_0}} \to 
1$ as $\sqrt{\tfrac{3}{2\varkappa}}\,g\rbr{\gamma-\gamma_0} \to \pm \infty$,
eq.\ \eqref{eq:hr-first-20} also coincides asymptotically with a classical trajectory,
with the same $p_\gamma$ as in eq.\ \eqref{eq:hr-first-15}. In contrast to 
eq.\ \eqref{eq:hr-first-10}, one has two distinct trajectories, which approach the 
same classical trajectory in the above-mentioned asymptotic region, while they
depart from the trajectory near the classical turning point. The result 
is plotted in fig.\ \ref{fig:linear-wp-rl-1der}.

Now consider a classical trajectory that is implicitly given by an equation
$\rfun{f}{x, y} = 0$. This works only in two dimensions; for $d$-dimensions, 
$d > 2$, one needs $d-1 > 1$ equations to specify an implicit curve. One can
intuitively imagine a wave-packet that \enquote{peaks around} this trajectory, 
the density of which is given by \cite[eq.\ (6.3)]{Halliwell2009}
\begin{align}
	\rho = \ee^{-f^2}\,,
	\label{eq:halliwell6.3-10}
\end{align}
so that the density $\rho$ peaks to $1$ at $f = 0$, and is less than $1$ for 
$f \neq 0$.

Using the first-derivative test with an arbitrary variable $x$, one has
\begin{align}
0 = \partial_x \rho = -2 \rho f\,\partial_x f\,,
\end{align}
and therefore
\begin{align}
f = 0\,,
\qquad\text{or}\qquad
\partial_x f = 0\,.
\end{align}
Hence the trajectory $f = 0$ is \emph{included} in the result of the
first-derivative test.

The first-derivative test is intuitive and easy to implement. However, it is not
covariant under coordinate transformation; moreover, one can construct examples 
where the test does not give sensible results, see fig.\ \ref{fig:heli-g-20}. One 
may imagine using the eigenvector field of the Hessian 
$\partial_i \partial_j \rho$ as the \enquote{principle directions} and perform
a directional derivative test with respect to them. This is the approach in 
\cite{Eberly1996}.

Unfortunately, the directional derivative test is not practical in higher
dimensions, where no generic expression for roots of the algebraic eigenvalue
equation exists. In addition, the smoothness of the eigenvector field is
difficult to establish. Moreover, upon moving to (pseudo-)Riemannian geometry,
one needs to deal with the $\rbr{1, 1}$-Hessian tensor, which is not 
symmetric as a matrix, and the analysis is lost in challenging calculations. 
We now move forward to the other two approaches of ridge-lines.

%123456789012345678901234567890123456789
\section{Classical predictions as contour ridge-lines}
\label{sec:hr-pg}
%1234567890123456789012345678901234567890123456789012345678901234567890123456789

In subsection \ref{ssec:hr-pg} we first describe the ridge-lines in terms of a certain
character of the contour lines. One can imagine finding the \emph{locally most 
curved neighbourhoods} on the contour lines, the trajectory of which forms a 
ridge- or dale-line. The defining equation of this approach was first written
down by Barr{\'e} de Saint-Venant in 1852 \cite{MdSV1852a} without derivation.
We refer to \cite{Eberly1996} for a comprehensive explanation.

We will begin with the two-dimensional Euclidean case, where there are two
equivalent definitions of the contour ridge-lines, both of which can be 
generalised to higher dimensions, as well as to (pseudo-)Riemannian geometry.
For the \enquote{linear} wave-packet in eq.\ \eqref{eq:linear-pac-Psi}, the
contour approach can directly be applied.

In this subsection \ref{sec:asp-con} we  establish a scenario with an exponential 
wave-packet, in which the contour approach gives intuitive results. We then
generalise this scenario with a slowly varying amplitude and show that an
intuitive result is still contained in the result. We show how the redundant
results can be identified with a toy example.

\subsection{The contour ridge-lines}
\label{ssec:hr-pg}

%123456789
\paragraph{First definition}

%1234567890123456789012345678901234567890123456789012345678901234567890123456789

In topography, contour lines give the altitude intrinsically. One can formulate
the ridge- and dale-lines in terms of the contour lines as follows
 \cite[sec.\ 4.1]{Markoski2018}:
\begin{quote}
	When representing ridges, contour lines are elongated towards ridge stretch
	and they are convex as they are turned towards the fall of the ridge or the
	ground \ldots
\end{quote}

Mathematically, one considers a $C^2$ real function $\rfun{\rho}{x, y}$, 
the contour lines $\gamma_c$ of which are given by the implicit equation 
$\rho \equiv c $. 

Having the idea of \enquote{locally most curved neighbourhoods} in the 
introduction in mind, now let $\rfun{\kappa}{x,y}$ be a \emph{characteristic
function}, such that the crossing of a ridge and the contour $\gamma_c$ is 
an extremum of $\kappa$ on $\gamma_c$. This gives the first definition of a 
contour ridge-line, namely \emph{the ridge-line is regarded as the locus of 
extrema of $\kappa$ under the constraint $\rho = c$}.

The statement can be formulated by the method of Lagrange multipliers,
\begin{subequations}
	\begin{align}
		\dd \rho &= \lambda_c\,\dd \kappa\,,
		\label{eq:lag-mul-10}\\
		\rho &= c\,,
	\end{align}
\end{subequations}
where $\lambda_c$ is the Lagrange multiplier. Equation \eqref{eq:lag-mul-10} can be 
separated into a system of equations in the bases $\dd x$ and $\dd y$.
Then eliminating $\lambda_c$ %in eq.\ \eqref{eq:lag-mul-10} 
gives
\begin{align}
	0 &= \rho_{,x} \kappa_{,y} - \rho_{,y} \kappa_{,x}\,,
	\label{eq:ridge-line-gen-05}
\end{align}
where \enquote{$_{,}$} denotes partial derivative 
\cite[eq.\ (2.25)]{Misner2018}.

In practice, one can use the squared norm  
of $\dd \rho$ as the characteristic function
\begin{align}
	\kappa = \rfun{\kappa_\text{sqr}}{x,y} = \rho_{,x}^2 + \rho_{,y}^2\,.
	\label{eq:square-cha-fun}
\end{align}
Substituting eq.\ \eqref{eq:square-cha-fun} in eq.\ \eqref{eq:ridge-line-gen-05} results in 
the \emph{de Saint-Venant equation for ridges} (dSVr) 
\cite{MdSV1852a,Koenderink1993}
\begin{align}
	0 &= \rho_{,x}\rho_{,y} \rbr{\rho_{,x,x}-\rho_{,y,y}}
		- \rbr{\rho_{,x}^2 - \rho_{,y}^2} \rho_{,x,y}\,.
	\label{eq:dSVr-10}
\end{align}

%123456789
\paragraph{Second definition}

%1234567890123456789012345678901234567890123456789012345678901234567890123456789

To see the mathematical structure more clearly, we use the generalisation of
eq.\ \eqref{eq:square-cha-fun} in eq.\ \eqref{eq:gamma-20}. Substituting the latter in
eq.\ \eqref{eq:lag-mul-10} gives the tensorial equation
\begin{align}
	\rho_{;i}=2\lambda_c \rho_{;i}{}^{;j} \rho_{;j}\,.
	\label{eq:hr-pg-eig-10}
\end{align}
In other words, $\rho^{;i}$ is an eigenvector of its Hessian 
$\rho^{;i}{}_{;j}$. This gives the second characteristic of a contour 
ridge-line: it is \emph{the locus of points where the gradient is an 
eigenvector of the Hessian}.

%123456789
\paragraph{Generalisations}

%1234567890123456789012345678901234567890123456789012345678901234567890123456789

The results above in two dimensions can easily be generalised to higher 
dimensional (pseudo-)Riemannian spaces. From eq.\ \eqref{eq:lag-mul-10} one can derive
\begin{align}
	0 = \dd \rho \wedge \dd \kappa\,,
	\label{eq:ridge-line-gen-10}
\end{align}
which takes the place of eq.\ \eqref{eq:ridge-line-gen-05}. For 
eq.\ \eqref{eq:square-cha-fun}, the generic version reads
\begin{align}
	\kappa_\text{sqr} &= \star^{-1}\rbr{\dd \rho\wedge\star\dd \rho}
	= \dd \rho^\sharp \intprod \dd \rho = g^{ij} \rho_{;i} \rho_{;j}\,,
	\label{eq:gamma-20}
\end{align}
where $\star$ is the Hodge star operator \cite[sec.\ 28]{Burke1985}, $^\sharp$ 
is a musical isomorphism, $\intprod$ is the interior product or contraction 
\cite[sec.\ 23]{Burke1985}, $g^{ij}$ is the inverse metric, and the symbol 
$_{;}$ denotes the covariant derivative with respect to an affine connection 
\cite[sec.\ 85]{Landau2en}. 

Inserting eq.\ \eqref{eq:gamma-20} in eq.\ \eqref{eq:ridge-line-gen-10} gives the 
\emph{covariant dSVr equation}
\begin{align}
	0 &= \dd \rho \wedge \dd \rbr{\dd \rho^\sharp \intprod \dd \rho}\,.
	\label{eq:dSVr-20}
\end{align}
This equation is to be understood as imposing all its components to be zero, 
and therefore defining an \emph{implicit} curve.

%1234567890
\paragraph{Application to the \enquote{linear} wave-packet}

%1234567890123456789012345678901234567890123456789012345678901234567890123456789

The contour approach can immediately be applied to the \enquote{linear} 
wave-packet in eq.\ \eqref{eq:linear-pac-Psi}. Using the DeWitt metric
in eq.\ \eqref{eq:prototype-11}, the de Saint-Venant equations
for ridges \eqref{eq:dSVr-20} can be factorised such that
\begin{subequations}
\begin{align}
0 &= y\,,\qquad \text{or}
\\
0 &= x^3 \rfun{\sinh}{y}^4
	- x^2 \rfun{\cosh}{y} \rfun{\sinh}{y}^2
	- x \rfun{\cosh}{y}^2
	+ \rfun{\cosh}{y}\,,
\label{eq:dSVr-linear-10}
\end{align}
\end{subequations}
where $x > 0$ is given in eq.\ \eqref{eq:transform-20},
$y = \sqrt{\frac{3}{2\varkappa}}\, g\rbr{\gamma-\gamma_0}$. One can solve $x$ 
from eq.\ \eqref{eq:dSVr-linear-10} in terms of $y$,
\begin{align}
\begin{split}
3x_k &= 1+4\cos\frac{2k\pp
		+\sfun{\arctan}{19-8\rfun{\cosh}{2y},
			3\sqrt{48\rfun{\cosh}{2y}-33}}}{3}\,,
\\
k &= 0, 1, 2\,,
\end{split}
\label{eq:hr-pg-linear-10}
\end{align}
where $\rfun{\arctan}{x, y}$ gives $\varphi \in [0, 2\pp)$ such that
$\cos\varphi = \frac{x}{\sqrt{x^2 + y^2}}$,
$\sin\varphi = \frac{y}{\sqrt{x^2 + y^2}}$.

In eq.\ \eqref{eq:hr-pg-linear-10}, since 
\begin{align}
\lim_{y\to\infty} \sfun{\arctan}{
		19-8\rfun{\cosh}{2y}, 3\sqrt{48\rfun{\cosh}{2y}-33}}
	= \pp\,,
\end{align}
one obtains
\begin{align}
\lim_{y\to\infty} x_k = \rbr{-}^{k+1} 3\,.
\end{align}
Therefore, the cases $k = 0$ and $2$ give positive $x$ and real $\chi$ as 
$\chi \to \infty$, whereas $k = 1$ does not. Exact calculation shows that 
$x_2 < 0$ for all $y \in \BbbR$, and is to be excluded.

\begin{figure}
\begin{center}
\begin{subfigure}{.49\linewidth}
	\begin{center}
\includegraphics{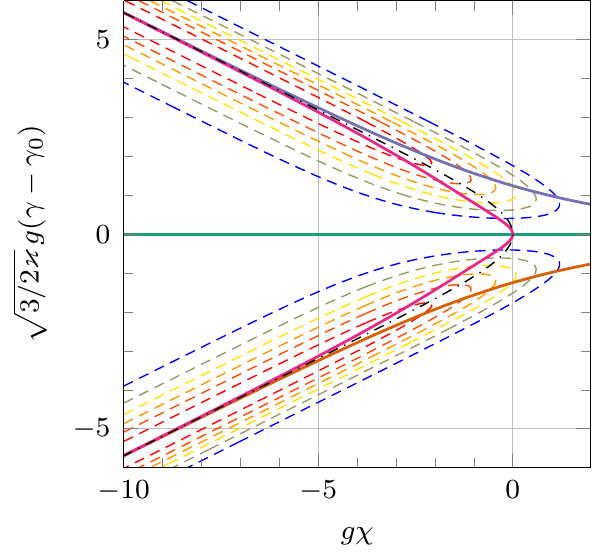}
	\end{center}
	\end{subfigure}
\caption%
	{The contour approach shown with the \enquote{linear} wave-packet in
	eq.\ \eqref{eq:linear-pac-Psi}. We have used the \emph{Lorentzian} minisuperspace metric in
	eq.\ \eqref{eq:prototype-11}.
	The orange and 
	purple lines are the results from first-derivative test, whereas the dark dash-dotted line is classical solution
	eq.\ \eqref{eq:hr-first-10}.
	The pink solid line is derived by the contour 
	approach in eq.\ \eqref{eq:hr-pg-linear-10}.
	\label{fig:linear-wp-rl-dSVr}}
\end{center}
\end{figure}

These results are plotted in fig.\ \ref{fig:linear-wp-rl-dSVr}. One sees a redundant
line $y = 0$ that is a dale, a pink line that resembles a classical trajectory,
and two further solid lines that converge to the same classical trajectory as
$\gamma \to \pm\infty$.
Nevertheless, the deviation from classical trajectory is apparent. 
More precisely, the classical prediction from the ridge-lines by the contour approach does not match the classical trajectory around the turning point,
which can be regarded as quantum correction to the classical theory.

%1234567890
\paragraph{Curvature as the characteristic function}

%1234567890123456789012345678901234567890123456789012345678901234567890123456789

In two dimensions, it is tempting and intuitive to use the curvature of the
contours as the characteristic function. We argue that this choice will not fit
our purpose. Upon generalising to higher dimensions, the curvature of an 
$\rbr{n-1}$-dimensional contour becomes the scalar-valued
second fundamental form, 
which is a symmetric tensor.
One may want to further analyse this tensor, and study its orthonormal
eigenvectors \cite{Eberly1996}.

Unfortunately, for the cases where the (DeWitt) metric is indefinite (e.g.\
Lorentzian), the second fundamental form is defined differently for the time- 
and space-like patches \cite[sec.\ 1.2.4]{Anciaux2010a}, which discontinues 
at the null edge, where the second fundamental form is again defined
differently \cite{Kupeli1987}. The reason is that, for time- and space-like 
hypersurfaces, the second fundamental form depends on the
choice of a unit normal vector, which of course discontinues going from a 
time-like patch to a space-like patch. Moreover, the eigenvectors of the
second fundamental form may also not exist 
(\cite[sec.\ 2.5.(2)]{Anciaux2010a}).

The contour ridge-line is based on first- and second-derivatives of $\rho$ and 
always give equations for an algebraic curve.
However, aside from sensible 
ridge-lines, this approach also gives counter-intuitive curves. 

%123456789012345678901234567890123456789
\subsection{Aspects of the contour approach}
\label{sec:asp-con}
%1234567890123456789012345678901234567890123456789012345678901234567890123456789

%123456789
\paragraph{Invariance under regular transformation and applications}

%1234567890123456789012345678901234567890123456789012345678901234567890123456789

For a transformation $\rho \to F \circ \rho$, the dSVr equation 
\eqref{eq:dSVr-20} transforms to
\begin{align}
	0 = \rbr{\frde{F}{\rho}}^3 \dd \rho \wedge
	\dd \rbr{\dd \rho^\sharp \intprod \dd \rho}\,.
	\label{eq:property-10}
\end{align}
If $F$ is strictly monotonic, i.e.\ $\dd F/\dd\rho\neq 0$, the extra factor is 
non-zero, and eq.\ \eqref{eq:property-10} gives the same ridge-line as 
eq.\ \eqref{eq:dSVr-20}.

Now we move back to the two-dimensional wave-packet eq.\ \eqref{eq:halliwell6.3-10}.
Since $\ee^{x}$ increases monotonically with respect to $x$, applying the 
above-mentioned property gives the ridge-line
\begin{align}
	0 &= \dd \rbr{f^2} \wedge \dd
	\rbr{\dd \rbr{f^2}^\sharp \intprod \dd \rbr{f^2}}
	\nonumber \\
	&= 8f^3\,\dd f\wedge \dd 
	\rbr{\dd f^\sharp \intprod \dd f}\,,
\end{align}
which means
\begin{subequations}
	\begin{align}
		0 &= f \qquad \text{or}
		\label{eq:property-20} \\
		0 &= \dd f\wedge\dd \rbr{\dd f^\sharp \intprod \dd f}\,.
		\label{eq:property-30}
	\end{align}
\end{subequations}
Equation \eqref{eq:property-20} gives what we wanted to set up, whereas 
eq.\ \eqref{eq:property-30} gives the ridge- (or dale-)line of $f$ itself.

\begin{figure}
	\begin{center}
	\begin{subfigure}{.49\linewidth}
	\begin{center}
		\includegraphics{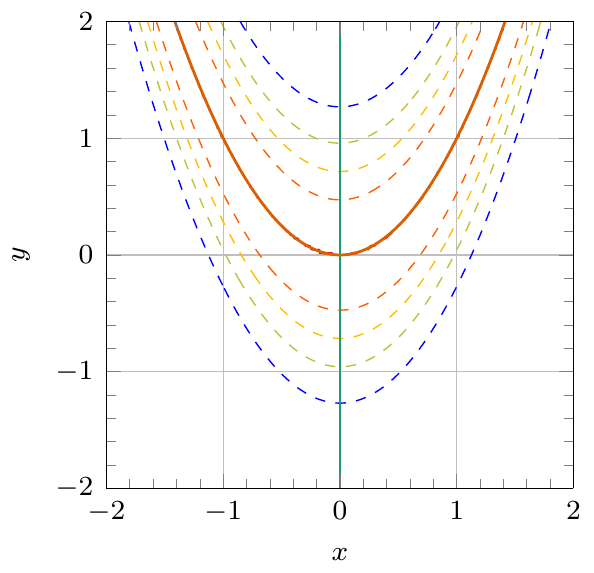}
	\end{center}
	\end{subfigure}
	\end{center}
	\caption[$\rho=\ee^{-(y-x^2)^2}$ and the contour ridge-lines]{%
	Density function $\rho=\ee^{-(y-x^2)^2}$ and the contour ridge-lines, 
	which are $x = 0$ (green, ridge-line for $y < 0$ and dale-line for $y > 0$)
	and $y = x^2$ (orange). Incidentally, these lines are also given by the
	first-derivative test $\rho_{,x}=0$, $\rho_{,y}=0$.
	\label{fig:basic-10}}
\end{figure}

This is easier to see with the toy example
\begin{align}
\rfun{f}{x,y} = y-x^2\,,
\label{eq:fyx2}
\end{align}
so that $f = 0$ gives the parabola $y = x^2$. There is an additional
solution to the dSVr equation, $x=0$, satisfying eq.\ \eqref{eq:property-30}. 
See fig.\ \ref{fig:basic-10}.

The parabola $y = x^2$ is what we wanted. However, we also get $x = 0$,
which is a dale-line for the density function $f = y - x^2$; as for 
$\rho = \ee^{-f^2}$, it is a ridge-line for $y < 0$, and a dale line for 
$y > 0$. This line is a concrete mathematical result, although it does not fit
our expectation.

%123456789
\paragraph{Modulation and redundant lines}
%123456789

The results for the wave-packet in eq.\ \eqref{eq:halliwell6.3-10} can be 
generalised to the narrow wave-packet with varying amplitude
\begin{equation}
\rfun{\rho}{x,y} = \rfun{g}{x,y}\ee^{-\frac{\rfun{f}{x,y}^2}{2\sigma^2}}\,,
\label{eq:halliwell6.3-20}
\end{equation}
where $\sigma$ is a constant, $\sigma \ll \vbr{\nabla g}$ characterising the
narrowness, and $g$ is a modulation. Substituting eq.\ \eqref{eq:halliwell6.3-20} 
into eq.\ \eqref{eq:dSVr-10} gives
\begin{equation}
0 = g^3 f^3 \rbr{p_f q_f \rbr{r_f-t_f} -
	\rbr{p_f^2 - q_f^2} s_f} + \rfun{O}{\sigma^2}\,,
\label{eq:dSVr-30}
\end{equation}
where $\rbr{p_f, \ldots, t_f}$ are the symbols with respect to $f$.

As $\sigma\to 0^+$, the wave-packet becomes sharper and sharper; except for an
additional factor $g^3$, the leading-order dSVr equation recovers the case 
without modulation. At the limit $\sigma=0^+$, the wave-packet becomes a wall 
with zero width, and extends along the classical trajectory $f=0$. 
Equation \eqref{eq:dSVr-30} shows that a slow modulation does not drastically
change the ridge-lines.

The narrow WKB Gaussian wave-packets in section \ref{ssec:packet-gaussian} is an
instance of this model. The heuristic arguments we used in that section can now
be replaced with the derivation in eq.\ \eqref{eq:dSVr-30}.

\begin{figure}
	\begin{subfigure}{.49\linewidth}
	\begin{center}
		\includegraphics{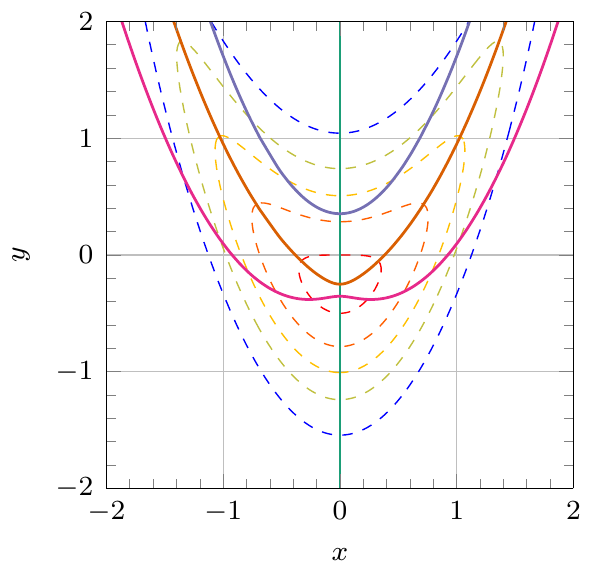}
	\end{center}
		\caption{$\rho = \ee^{-(y-x^2)^2-y/2}$}
		\label{fig:basic-20-2}
	\end{subfigure}
	\begin{subfigure}{.49\linewidth}
	\begin{center}
		\includegraphics{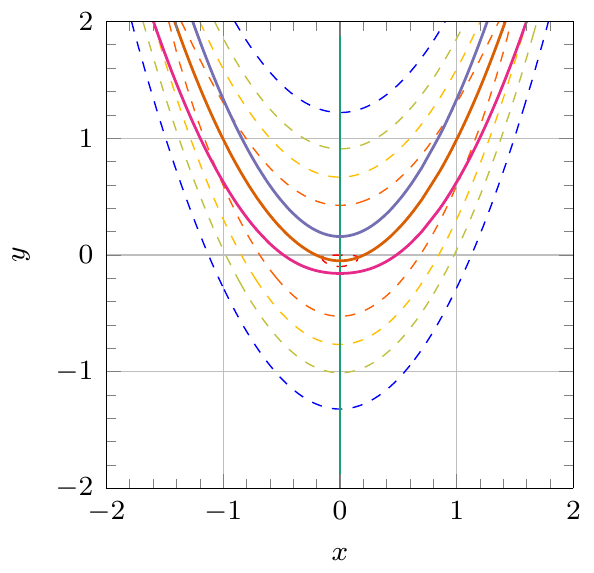}
	\end{center}
		\caption{$\rho = \ee^{-(y-x^2)^2-y/10}$}
		\label{fig:basic-20-10}
	\end{subfigure}
	\caption[$\rho=g\,\ee^{-(y-x^2)^2}$ and the Euclidean contour ridge-lines]{%
	Density function $\rho=g\,\ee^{-(y-x^2)^2}$ and the contour ridge-lines for
	$g = \ee^{-y/2}$ and $g = \ee^{-y/10}$ with a Euclidean metric. The green 
	line $x = 0$ and the orange line are (qualitatively) the same as in
	fig.\ \ref{fig:basic-10}; however, the dSVr equation \eqref{eq:dSVr-20} also 
	gives the purple and the pink lines, which are apparently neither ridge- 
	nor	dale-lines.
	\label{fig:basic-20}}
\end{figure}

Exact calculation reveals that the approximation we used to derive 
eq.\ \eqref{eq:dSVr-30} loses details. To see this, we also modulate eq.\
\eqref{eq:fyx2} by
\begin{align}
g = \ee^{-2\epsilon y}\,,
\qquad \epsilon = \frac{1}{2}, \frac{1}{10}\,.
\label{eq:gexpf2-g}
\end{align}
The dSVr equation for $\rho = g\,\ee^{-(y-x^2)^2}$ with $g$ given in 
eq.\ \eqref{eq:gexpf2-g} reads
\begin{align}
\begin{split}
0 = 16x &[-2 y^3 + 2 y^2 \rbr{3 x^2 - \epsilon}
	+y \rbr{-6 x^4 + 8 \epsilon x^2 + \epsilon} \\
	&
	+2 x^6 - 6 \epsilon x^4 - \epsilon x^2 + \epsilon^2]\,,
\end{split}
\end{align}
which has been factorised into $x = 0$, and a term cubic in $y$. One can solve 
$y$ in terms of $x$ from the factor in a square bracket, where the three roots 
$y = \rfun{y}{x}$ are all real. See fig.\ \ref{fig:basic-20}.

Only one of the three roots approaches $y = x^2$ as $x \to \infty$. This can 
be seen by expanding $\rfun{y}{x}-x^2$ at $\epsilon = 0^+$, which yields
\begin{subequations}
\begin{align}
\rfun{y_{1, 2}}{x} - x^2 &=
	\mp \sqrt{\frac{1+4x^2}{2}} \epsilon^{1/2} + 
	\rbr{-1+\frac{1}{1+4x^2}}\frac{\epsilon}{2}
	+ \rfun{\Omicron}{\epsilon^{3/2}}\,,
\label{eq:gexpf2-dSVr-root-12}\\
\rfun{y_{3}}{x} - x^2 &= -\epsilon\frac{1}{1+4x^2}
	+ \rfun{\Omicron}{\epsilon^{2}}\,.
\label{eq:gexpf2-dSVr-root-3}
\end{align}
\end{subequations}
As $x \to \pm\infty$, $y_3 - x^2$ converges to $0$, whereas $y_{1,2} - x^2$ 
diverge, and can be interpreted as the locus of the \enquote{locally flattest
places on the contour}, resembling $x=0$ for $y = x^3$. 

The extra curves in eq.\ \eqref{eq:gexpf2-dSVr-root-12} seem to be a common 
feature of the dSVr equation. Here we have managed to remove them by 
asymptotic analysis at infinity, recovering the intuitive result $y_3$.
The extra line $x = 0$ has been discussed at the end of the last part.

%123456789
\paragraph{Two-dimensional hydrogen atom revisited}

%123456789

\begin{figure}
	\centering
	\begin{subfigure}{.49\linewidth}
		\centering
		\includegraphics{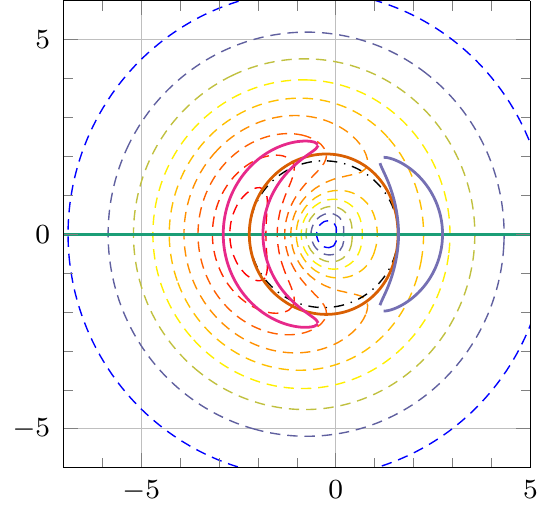}
	\end{subfigure}
\caption[Stationary wave-packet of the two-dimensional hydrogen atom 2]%
{Stationary wave-packet
$\vbr{\rfun{\varPsi_{1, \frac{23}{24}}}{\xi, \varphi}}^2$ 
of the two-dimensional hydrogen atom with $n=1$, $q=\frac{23}{24}$. See 
section \ref{sec:sta-wp} for details.
The thick lines with colour are solutions of the dSVr equation, whereas the 
dash-dotted line is the \enquote{best-fit trajectory} that crosses the maxima, 
adapted from the orange line in fig.\ \ref{fig:2dhyd-sta-pac-10-20}. The 
discontinuities within the same color are a numerical artefact.
\label{fig:2dhyd-sta-pac-25}}
\end{figure}

The binomial stationary wave-packets of two-dimensional hydrogen atom,
described in section \ref{sec:sta-wp}, can also be studied by the contour approach.
For $n = 1$, the dSVr equation is a sextic equation with respect to the 
dimensionless radial coordinate $\xi$, which has a quadratic and a quartic 
factor
\begin{subequations}
\begin{align}
\begin{split}
0 &= -x^2 \rbr{\sqrt{\ldots} \cos\varphi-1}^2
	\\
	&\quad\,
	+ x \left(\sqrt{\ldots}^2 \rfun{\cos}{2\varphi}+3 \sqrt{\ldots}^2
		-6 \sqrt{\ldots} \cos\varphi+2\right)
	\\
	&\quad\,
	+\sqrt{\ldots} \rbr{2 \cos\varphi-3 \sqrt{\ldots}}\,,
	\qquad\text{or}
\end{split}
\\
\begin{split}
0 &= +4 x^4 \rbr{\sqrt{\ldots} \cos\varphi-1}^3
	\\
	&\quad\,
	+ x^3 \Big[-4 \sqrt{\ldots}^3 \rfun{\cos}{3\varphi}
		+30 \sqrt{\ldots}^2 \rfun{\cos}{2\varphi}
		\\
		&\qquad\quad\quad
		-4 \left(5 \sqrt{\ldots}^2+16\right) \sqrt{\ldots} \cos\varphi
		+38 \sqrt{\ldots}^2+20\Big]
	\\
	&\quad\,
	+x^2 \Big\{6 \sqrt{\ldots}
		\Big[+2 \left(3 \sqrt{\ldots}^2+8\right) \cos\varphi
			\\
			&\qquad\qquad\qquad\ \ 
			-5 \sqrt{\ldots} \rfun{\cos}{2\varphi}\Big]
		-74 \sqrt{\ldots}^2-28\Big\}
	\\
	&\quad\,
	+4 x \left[-2 \left(3 \sqrt{\ldots}^2+4\right) \sqrt{\ldots} \cos\varphi
		+13 \sqrt{\ldots}^2+2\right]
		\\
		&\quad\,
	-12 \sqrt{\ldots}^2\,,
\end{split}
\end{align}
\end{subequations}
where $\sqrt{\ldots} \coloneqq \sqrt{1-q^2}$.
We are therefore able to obtain solutions in terms of roots. Aside from 
$\sin\varphi = 0$, there are six solution $\xi = \rfun{\xi}{\varphi}$, three in 
which are real and positive near $\varphi = 0$ and $\varphi = \pp$; one is from 
the quadratic factor and has a simple form, while the other two are very 
complicated. We managed to plot them in fig.\ \ref{fig:2dhyd-sta-pac-25}. 

One sees that the orange ridge given by the dSVr equation is very close
to the \enquote{best-fit trajectory} that passes through the maxima of the
wave-packet. Like in the case $\rho = g\,\ee^{-f^2}$, there are two 
additional lines, which might be the locally flattest points of the contours.

%123456789
\paragraph{Lorentzian signature}

%123456789

In quantum cosmology, the minisuperspace DeWitt metric usually has a Lorentzian
signature. For the Lorentzian metric
\begin{align}
\dd s^2 = -\dd t^2 + \dd x^2\,,
\label{eq:gexpf2-lor-met}
\end{align}
the Lorentzian dSVr, according to eq.\ \eqref{eq:dSVr-20}, reads
\begin{align}
	0 &= - \rho_{,x}\rho_{,t} \rbr{\rho_{,x,x}+\rho_{,t,t}}
		+ \rbr{\rho_{,x}^2 + \rho_{,t}^2} \rho_{,x,t}\,.
	\label{eq:dSVr-90}
\end{align}
In fig.\ \ref{fig:linear-wp-rl-dSVr}, we have already shown a sensible
result with contour ridge-lines in a Lorentzian signature.

\begin{figure}
	\begin{subfigure}{.49\linewidth}
	\begin{center}
		\includegraphics{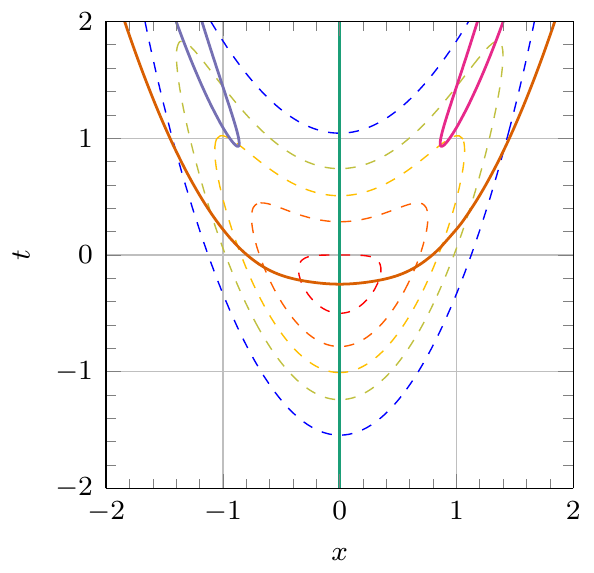}
	\end{center}
		\caption{$\rho = \ee^{-(t-x^2)^2-t/2}$}
		\label{fig:basic-30-2}
	\end{subfigure}
	\begin{subfigure}{.49\linewidth}
	\begin{center}
		\includegraphics{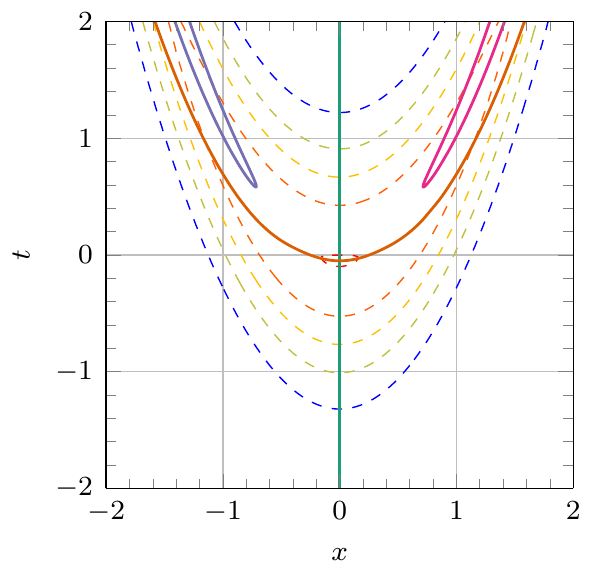}
	\end{center}
		\caption{$\rho = \ee^{-(t-x^2)^2-t/10}$}
		\label{fig:basic-30-10}
	\end{subfigure}
	\caption[$\rho=g\,\ee^{-(t-x^2)^2}$ and the Lorentzian contour 
	ridge-lines]{Density function $\rho=g\,\ee^{-(t-x^2)^2}$ and the 
	contour ridge-lines for
	$g = \ee^{-t/2}$ and $g = \ee^{-t/10}$ with an Lorentzian metric
	\eqref{eq:gexpf2-lor-met}. The green line $x = 0$ is the same as in
	figs.\ \ref{fig:basic-10} and \ref{fig:basic-20}. The orange line fits intuition better
	near $x = 0$, whereas the other two lines both have a sharp turning point, 
	and one of the branches fits the intuitive ridge in the asymptotic 
	region.}
	\label{fig:basic-30}
\end{figure}

For the $\rho = g\,\ee^{-f^2}$ model, we can also mimic the scenario by 
replacing $y \to t$ in eqs.\ \eqref{eq:fyx2} and \eqref{eq:gexpf2-g}, and using the metric
in eq.\ \eqref{eq:gexpf2-lor-met}. The result can still be factorised to $x = 0$
and a cubic algebraic equation with respect to $t$, see 
fig.\ \ref{fig:basic-30}. 

Intriguingly, \emph{none} of the three curves given by the latter factor lies 
on the intuitive ridge globally; instead, for the turning and asymptotic 
regions, there is one branch for each case that fits well with intuition.

%123456789012345678901234567890123456789
\section{Classical predictions as stream ridge-lines}
\label{sec:hr-sg}
%1234567890123456789012345678901234567890123456789012345678901234567890123456789

Now we consider the ridges in terms of \emph{singular} stream-lines of the 
gradient vector field, which dates back to Rudolf Rothe in 1915 
\cite{rothe1915}. Heuristically, one imagines that water slowly flows from the 
top of a hill along the stream-lines of the gradient vector field. The water 
stream diverges from a ridge and converges to a dale. This is the intuitive 
notion of the singularity of the stream-lines along ridge- and dale-lines.

The stream approach is also adapted by modern computer scientists in image
processing and computer vision \cite{Koenderink1993,Koenderink1994}. The 
mathematics behind this approach is the inverse integral factor and inverse 
Jacobi multiplier, which work for two- and higher-dimensional cases, 
respectively \cite{Garcia2010,Berrone2003}. We will focus on the 
two-dimensional case.

After a general discussion in section
\ref{ssec:hr-sg},
we examine two families of density function,
for which the stream ridge-lines can be exactly solved in section \ref{sec:asp-str}. We then show that the
toy model $\rho = \ee^{-\rbr{y-x^2}^2}$ introduced in 
section \ref{sec:asp-con} belongs to one of the families. In the end we investigate 
the cases with a Lorentzian metric signature.

%123456789012345678901234567890123456789
\subsection{The stream ridge-lines}
\label{ssec:hr-sg}
%1234567890123456789012345678901234567890123456789012345678901234567890123456789

%123456789
\paragraph{Inverse integral factor}

%1234567890123456789012345678901234567890123456789012345678901234567890123456789

In $\mathbb{R}^2$ with Cartesian coordinates $\rbr{x, y}$, the contours of 
$\rho$ are defined by $\dd \rho = 0$, or $\rho = c$; dual to them are the 
stream-lines, characterised by $\dd w = 0$ or $w = c$, where
\begin{align}
	\theta\,\dd w = \star \dd \rho = -\rho_{,y}\,\dd x + \rho_{,x}\,\dd y\,,
	\label{eq:stream-rid-lin-05}
\end{align}
in which $\theta$ compensates the non-integrability of the right-hand side
and is therefore called an \emph{inverse integral factor}. One also has
\begin{align}
	0 = \rho_{,x} w_{,x} + \rho_{,y} w_{,y} =
		\star^{-1}\rbr{\dd \rho \wedge \star \dd w}\,.
	\label{eq:stream-rid-lin-07}
\end{align}
$\rbr{\theta, w}$ is unique up to
\begin{align}
\theta \to \theta/\rfun{F'}{w}\,,\qquad
w \to \rfun{F}{w}\,,
\label{eq:theta-w-f}
\end{align}
where $\rfun{F}{w}$ is an arbitrary function. One may worry that this
arbitrariness renders the stream approach not giving definite results, which
fortunately does not seem to be the case, see section \ref{sec:hr-dif}.

One sees that if $\theta = 0$ and $\rho_{,x} \neq 0 \neq \rho_{,y}$ at 
$\rbr{x_0, y_0}$, $w$ cannot be expanded by the Taylor theorem at 
$\rbr{x_0, y_0}$, since the linear term blows up by eq.\ \eqref{eq:stream-rid-lin-05} 
\cite[sec.\ 7]{rothe1915}. $\rbr{x_0, y_0}$ is said to be on a 
\emph{singular stream-line}.
%\cite{Hamburger1893a}.

One can imagine that if the ridge- and dale-lines are required also to be 
stream-lines themselves, then the neighbouring stream-lines converge to the 
former, and diverge from the latter along the direction of the gradient vector 
field. In other words, \emph{stream ridge- and dale-lines are singular 
stream-lines}. It has been shown that along these stream-lines, one has
\cite{Garcia2010}
\begin{subequations}
	\begin{align}
		\rfun{\theta}{x, y} = 0\,.
		\label{eq:stream-rid-lin-10}
\end{align}
The integrability condition $\dd \wedge \dd w = 0$, or 
$\theta_{,x,y} = \theta_{,y,x}$, gives the differential equation for $\theta$,
\begin{align}
	\rho_{,x} \theta_{,x} + \rho_{,y} \theta_{,y} =
	\rbr{\rho_{,x,x} + \rho_{,y,y}}\theta\,.
	\label{eq:stream-rid-lin-20}
\end{align}
\end{subequations}
Equations \eqref{eq:stream-rid-lin-10} and \eqref{eq:stream-rid-lin-20} define the stream ridge- and 
dale-lines.

%123456789
\paragraph{Generalisations}

%1234567890123456789012345678901234567890123456789012345678901234567890123456789

The results above in two dimensions can readily be generalised to 
$n$-dimensional curved spaces. Consider local coordinates 
$\rbr{x^1, \ldots, x^n}$, $n \ge 2$. The gradient vector field $v$ of $\rho$ is 
given by
\begin{align}
	v^i\,\partial_i \equiv v = \dd \rho^\sharp \coloneqq g^{ij} f_{,j} 
		\partial_i\,.
\end{align}
One has $\rbr{n-1}$ linearly independent $w$'s for the stream-lines, satisfying
\begin{align}
	0 = v^i\, \partial_{i} w = \rfun{v}{w}\,,
	\label{eq:stream-rid-lin-30}
\end{align}
which is the generalisation of eq.\ \eqref{eq:stream-rid-lin-07}. 
They are nothing else but the $\rbr{n-1}$ first integrals \cite{ArnoldODEen}, 
that require $\rbr{n-1}$ inverse integral factors $\theta$.

Similar to eq.\ \eqref{eq:stream-rid-lin-05}, one has for instance
\begin{align}
	\theta \,\dd w &= v^1\,\dd x^j - v^j\,\dd x^1\,, \qquad
	2 \le j \le n\,,
\end{align}
given $v^i \neq 0, 1 \le i \le n$.
All of the $\theta$'s satisfying the linear, first-order partial differential 
equation
\begin{align}
	v^i \theta_{,i} = \theta v^i{}_{;i}\,,
	\qquad\text{or} \qquad
	v\intprod\dd \theta = \theta\,\dd^{\dagger} v^{\flat} \,,
	\label{eq:pg-theta-30}
\end{align}
where $\dd^\dagger$ is the codifferential or the adjoint 
\cite[sec.\ 29]{Burke1985}. The solutions to eq.\ \eqref{eq:pg-theta-30} are called
\emph{inverse Jacobi multipliers} \cite{Berrone2003}, first appeared in 
\cite{Jacobi1844a}.

For Riemannian geometry, the stream approach seems to always give sensible 
results, in contrast with the contour approach and the simple first-derivative 
test. However, the approach involves giving the \emph{general integral} 
\cite[sec.\ 3.1.2]{evans2010partial} of the partial differential equation 
\eqref{eq:stream-rid-lin-30} or \eqref{eq:pg-theta-30}, which is only possible 
in very limited cases. Moreover, Lorentzian geometry gives rise to 
counter-intuitive configurations of gradient \emph{vector} fields, where the 
time-like component of the gradient one-form fields is flipped. This leaves us 
problems that are yet to be solved. See sections \ref{sec:hr-dif} and
\ref{sec:asp-str}.

%123456789012345678901234567890123456789
\subsection{Aspects of the stream approach}
\label{sec:asp-str}
%1234567890123456789012345678901234567890123456789012345678901234567890123456789

%123456789
\paragraph{Stream ridge-lines of two function families}

%123456789

For density functions of the following two forms
\begin{subequations}
	\begin{align}
		\rfun{\rho}{u, v} &= \rfun{f}{\rfun{f^{u}}{u}+\rfun{f^{v}}{v}}\,,
		\label{eq:new-ex-100}\\
		\rfun{\rho}{u, v} &= \rfun{f}{\rfun{f^{u}}{u}\rfun{f^{v}}{v}}\,
		\label{eq:new-ex-200}
	\end{align}
\end{subequations}
with the metric
\begin{align}
\begin{split}
	\dd s_1^2 &= \rfun{h}{u, v}^2\rbr{\mitsansg\, \dd u^2 + \dd v^2}\,,
	\\
	\mitsansg &= \pm\,,\qquad
	\rfun{h}{u, v} > 0
\end{split}
	\label{eq:new-met-100}
\end{align}
the stream-lines of the gradient vector field can be exactly solved.
Note that for the Euclidean signature $\mitsansg = +$, 
eq.\ \eqref{eq:new-met-100} includes the
bipolar, Cartesian, elliptic and planar parabolic coordinates for the flat 
geometry, and the stereographic coordinates for the spherical geometry, so that 
it is quite comprehensive. The Hodge-stars of the coordinate differentials read
\begin{align}
\star \dd u = \mitsansg \, \dd v\,,
\qquad
\star \dd v = -\dd u\,;
\label{eq:new-star-100}
\end{align}
one therefore gets
\begin{align}
\theta\,\dd w = \star \dd \rho = -\rho_{,v}\,\dd u
	+ \mitsansg \rho_{,u}\,\dd v\,.
\end{align}

By using eqs.\ \eqref{eq:stream-rid-lin-07}, \eqref{eq:new-met-100} and \eqref{eq:new-star-100},
one obtains for eq.\ \eqref{eq:new-ex-100}
\begin{subequations}
\begin{align}
	w &= \rfun{F}{-\mitsansg
		\int^{u} \frac{\dd \mu}{\rfun{f^{u\prime}}{\mu}}
		+ \int^{v} \frac{\dd \nu}{\rfun{f^{v\prime}}{\nu}}}
	\label{eq:new-ex-110a} \\
	\theta &= -\frac{1}{F'} f' \rfun{f^{u\prime}}{u}\rfun{f^{v\prime}}{v}\,,
	\label{eq:new-ex-110b}
	\end{align}
\end{subequations}
% Contour approach of de Saint-Venant: lengthy. Both gives $f' = 0$; the stream
% approach gives $\rho_{,u}\rho_{,v} = 0$, whereas the contour approach gives
% $\cbr{\ldots} = 0$.
and for eq.\ \eqref{eq:new-ex-200}
\begin{subequations}
\begin{align}
	w &= \rfun{F}{-\mitsansg
		\int^{u} \frac{\dd \mu}{\rbr{\ln \rfun{f^{u}}{\mu}}^\prime}
		+ \int^{v} \frac{\dd \nu}{\rbr{\ln \rfun{f^{v}}{\nu}}^\prime}}
	\\
	\theta &= -\frac{1}{F'} f' \rfun{f^{u\prime}}{u}\rfun{f^{v\prime}}{v}
	\label{eq:new-ex-210b}
\end{align}
\end{subequations}
Curiously, both eqs.\ \eqref{eq:new-ex-110b} and
\eqref{eq:new-ex-210b} includes the result from
the first-derivative test, $\rho_{,u} = 0$ or $\rho_{,v} = 0$.

%123456789
\paragraph{Application to the toy model}

%123456789

\begin{figure}
\begin{center}
\begin{subfigure}{.49\textwidth}
\begin{center}
\includegraphics{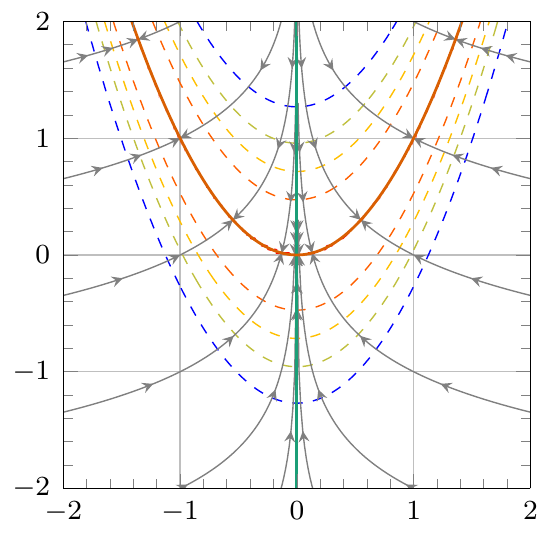}
\end{center}
\caption{Euclidean signature $\mitsansg = +$}
\end{subfigure}
\begin{subfigure}{.49\textwidth}
\begin{center}
\includegraphics{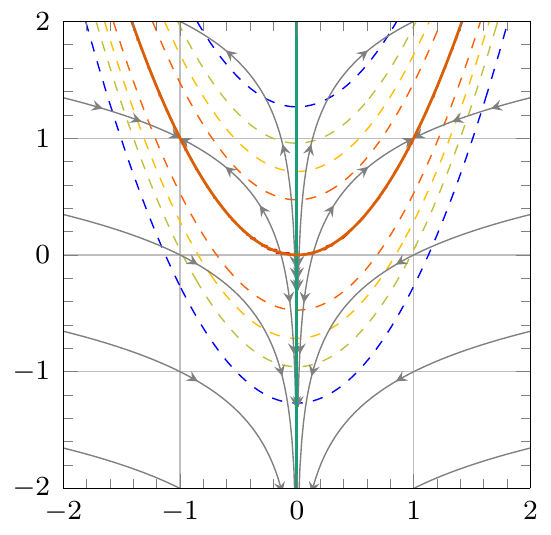}
\end{center}
\caption{Lorentzian  signature $\mitsansg = -$}
\end{subfigure}
\end{center}
\caption[$\rho = \ee^{-(y-x^2)^2}$ and the stream ridge-lines]%
	{Density function $\rho=\ee^{-(y-x^2)^2}$, the stream-lines of the
	gradient vector field in both Euclidean and Lorentzian geometry, and
	the stream ridge-lines which are $x = 0$ (green) and $y = x^2$ (orange).
	The Lorentzian results are to be understood with $y$ having the negative
	signature in the Minkowski metric.}
\label{fig:expf2sl}
\end{figure}

The toy model $\rho = \ee^{-\rbr{y-x^2}^2}$ in section \ref{sec:asp-con} has the
form of eq.\ \eqref{eq:new-ex-100}. One can adapt the results in 
eqs.\ \eqref{eq:new-ex-110a} and \eqref{eq:new-ex-110b} and get
\begin{subequations}
\begin{align}
	w &= \rfun{F}{\mitsansg y+\frac{1}{2}\ln x}\,,
	\label{eq:expf2-s-100a}\\
	\theta &= 4\mitsansg \ee^{-(y-x^2)^2} \frac{x\rbr{y-x^2}}%
	{\rfun{F'}{\mitsansg y+\frac{1}{2}\ln x}}\,.
	\label{eq:expf2-s-100b}
\end{align}
\end{subequations}
See fig.\ \ref{fig:expf2sl}. The Lorentzian results are to be understood
with $y$ having the negative signature in the Minkowski metric.
Equation \eqref{eq:expf2-s-100b} gives the same ridge-lines as 
in the contour approach, as well as in the first-derivative test, 
$y=x^2$ and $x=0$.

Now we move to the modulated toy model 
$\rho = \rfun{g}{x,y}\,\ee^{-\rbr{y-x^2}^2}$. 
Using $g_\epsilon=\ee^{-2\epsilon y}$, eq.\ \eqref{eq:heli-w-10} becomes
\begin{equation}
\mitsansg \rbr{-x^2+y+\epsilon} w_{,y} + 2 x \rbr{x^2-y} w_{,x}=0.
\end{equation}
For $\epsilon\ll 1$, one uses the series test solution
\begin{equation}
w=\sum_{n=0}^\infty w_n \epsilon^n\qquad \text{with}\qquad
w_0= \rfun{F}{\mitsansg y+\frac{1}{2}\ln x}\,,
\end{equation}
and for $n \ge 0$,
\begin{align}
\rbr{y-x^2}\rbr{2 \mitsansg x \partial_x w_{n+1} - \partial_y w_{n+1}} = 
	\partial_y w_{n}\,.
\end{align}
On the other hand, 
\begin{equation}
\theta = \mitsansg \frac{\rho_{, x}}{w_{, y}}
	= -\frac{\rho_{, y}}{w_{, x}},
\end{equation}
where $\rho$ can also be expanded with respect to $\epsilon$, i.e. 
\begin{equation}
\rho=\rho_0 \rbr{1+ \sum_{n=0}^{+\infty} \frac{\epsilon^n}{n!}
\left.\frpa{^n\ee^{-2\epsilon y}}{\epsilon^n}\right|_{\epsilon = 0}}\,,
	\qquad
\rho_0 =  \ee^{-(y-x^2)^2}.
\end{equation}
This implies that $\theta \propto \partial_x \rho_{0} \propto x \rbr{y-x^2}$.

\begin{figure}
\begin{center}
\begin{subfigure}{.49\textwidth}
\begin{center}
\includegraphics{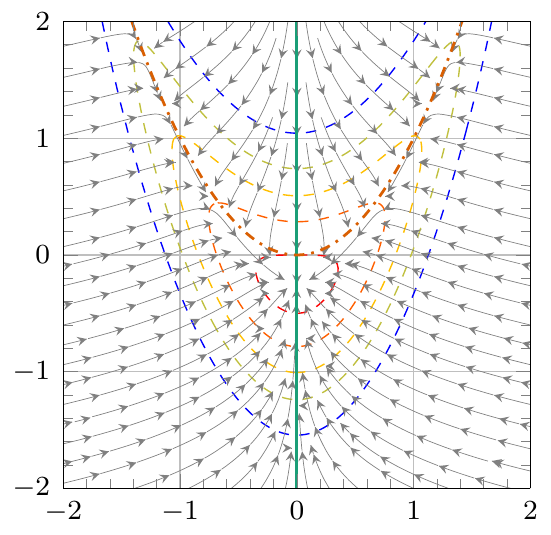}
\end{center}
\caption{$\rho = \ee^{-\rbr{y-x^2}^2-y/2}$}
\label{fig:expf2s-2}
\end{subfigure}
\begin{subfigure}{.49\textwidth}
\begin{center}
\includegraphics{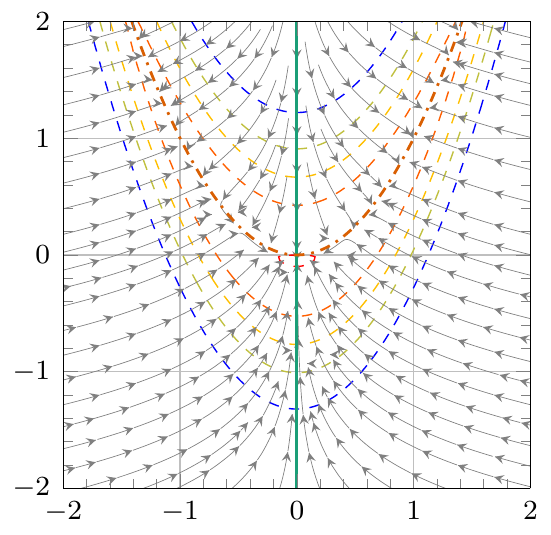}
\end{center}
\caption{$\rho = \ee^{-\rbr{y-x^2}^2-y/10}$}
\label{fig:expf2s-10}
\end{subfigure}
\begin{subfigure}{.49\textwidth}
\begin{center}
\includegraphics{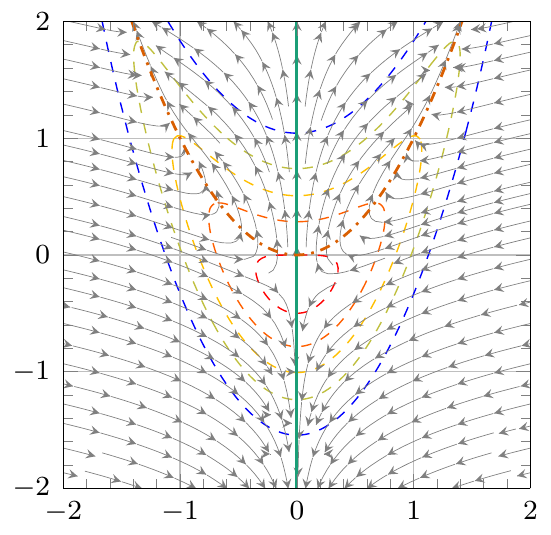}
\end{center}
\caption{$\rho = \ee^{-\rbr{t-x^2}^2-t/2}$}
\label{fig:expf2sl-2}
\end{subfigure}
\begin{subfigure}{.49\textwidth}
\begin{center}
\includegraphics{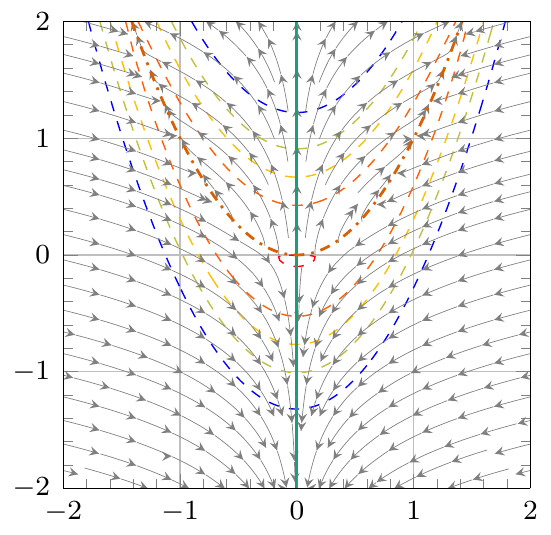}
\end{center}
\caption{$\rho = \ee^{-\rbr{t-x^2}^2-t/10}$}
\label{fig:expf2sl-10}
\end{subfigure}
\end{center}
\caption[$\rho=g\,\ee^{-(y-x^2)^2}$ and the stream ridge-lines]{%
	Density function $\rho=g\,\ee^{-(y-x^2)^2}$ and the contour ridge-lines 
	for $g = \ee^{-y/2}$ and $g = \ee^{-y/10}$ with the Euclidean and Lorentzian
	metrics. The green solid line $x = 0$ remains a ridge-dale-line, whereas
	the orange dash-dotted line is merely an approximation in the Euclidean
	case; the actual singular stream-lines seem to be under the orange lines. 
	The stream ridge-line in the Lorentzian signature is apparently more
	intriguing.}
\label{fig:expf2s}
\end{figure}

We failed to obtain a general integral $w$ for the modulated toy model
$\rho = g\,\ee^{-\rbr{y-x^2}^2}$. Numerically integrated stream-lines of the 
gradient vector field are plotted in fig.\ \ref{fig:expf2s}. One sees that for the 
Euclidean signature, the stream-lines indicate the fastest up-hill direction, 
in which the singular stream-lines are ridge- or dale-lines that fit the 
intuition. Moreover, the dash-dotted orange line $y = x^2$ is a good 
approximation of the actual ridge-line for small $\epsilon$ 
(fig.\ \ref{fig:expf2s-10}), but fails for larger $\epsilon$ (fig.\ \ref{fig:expf2s-2}); 
in other words, there are non-perturbative effects that cannot be revealed by 
the perturbative analysis above.

With the Lorentzian signature shown in figs.\ \ref{fig:expf2sl-2} and \ref{fig:expf2sl-10}, 
things become more complicated. The 
above-mentioned property, that the gradient vector field points to the
up-hill direction, is lost. Furthermore, the apparent ridge in the plot
is no longer accompanied by a possible singular stream-line; instead, on
the plot one sees a series of turning points that could play the role
of indicating a ridge-line that also fits human cognition.

\paragraph{Numerical applications to other models}

\begin{figure}
	\centering
	\begin{subfigure}{.49\linewidth}
		\centering
		\includegraphics{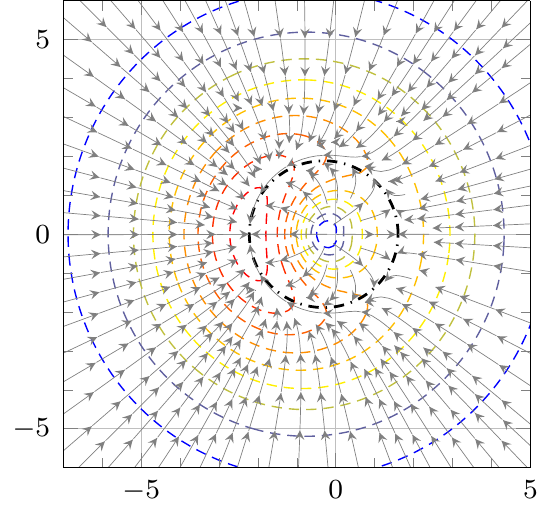}
		\caption{$2$d hydrogen atom $\varPsi_{1,\frac{23}{24}}$}
		\label{fig:2dhyd-sta-pac-30a}
	\end{subfigure}
	\begin{subfigure}{.49\linewidth}
		\centering
		\includegraphics{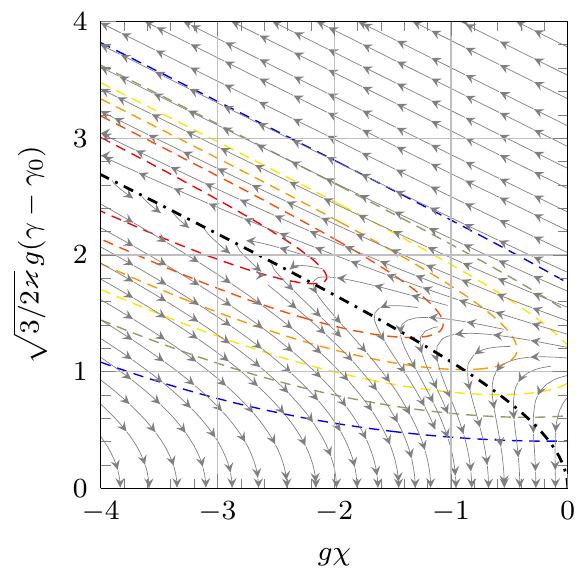}
		\caption{\enquote{Linear} wave-packet}
		\label{fig:2dhyd-sta-pac-30b}
	\end{subfigure}
\caption[Further numerical examples of stream ridge-lines]%
	{Numerical results of stream ridge-lines for the two-dimensional
	hydrogen atom \ref{fig:2dhyd-sta-pac-30a} and the \enquote{linear} 
	wave-packet \ref{fig:2dhyd-sta-pac-30b}. In fig.\ \ref{fig:2dhyd-sta-pac-30a},
	the geometry is Euclidean, and the dash-dotted line is the 
	\enquote{best-fit} classical elliptic trajectory, passing through the 
	maxima of the stationary wave-packet. In fig.\ \ref{fig:2dhyd-sta-pac-30b},
	the geometry is Lorentzian, and the dash-dotted line is the 
	\enquote{best-fit} trajectory used before.}
\label{fig:2dhyd-sta-pac-30}
\end{figure}

As mentioned before, the stream approach is difficult to obtain analytic 
results. For the two-dimensional hydrogen atom and the \enquote{linear}
wave-packet that were studied before, we make numeric plots of the stream-lines
of the gradient vector fields, see fig.\ \ref{fig:2dhyd-sta-pac-30}.

One sees again the good quality in the case with a Euclidean signature in
fig.\ \ref{fig:2dhyd-sta-pac-30a}, that no counter-intuitive lines are present.
There seems to be a singular stream-line that is very close to the 
\enquote{best-fit} classical trajectory. For the Lorentzian geometry,
the \enquote{best-fit} classical trajectory lies again near the 
\enquote{turning points} of the stream-lines, instead of being near
a singular stream-line.

%123456789012345678901234567890123456789
\section{Relations of the contour and stream approaches}
\label{sec:hr-dif}
%1234567890123456789012345678901234567890123456789012345678901234567890123456789

In this section \ref{sec:hr-dif} we compare the contour and stream approaches, as 
well as argue against the first-derivative test. Much of the material is
adapted from \cite{rothe1915,Koenderink1993}.

The contour and stream ridge-lines can be derived on the same footing.
In $\mathbb{R}^2$, from $\dd \rho = \rho_{,x} \,\dd x + \rho_{,y} \,\dd y$, 
eqs.\  \eqref{eq:square-cha-fun} and \eqref{eq:stream-rid-lin-05}, one deduces that 
\cite[sec.\ 5]{rothe1915}
\begin{align}
	\frac{1}{2}\,\dd \kappa_\text{sqr} &= R\,\dd \rho + \theta S\,\dd w\,,
\end{align}
where
\begin{subequations}
\begin{align}
	R &\coloneqq \frac{\rho_{,x}^2 \rho_{,x,x}
			+ 2 \rho_{,x} \rho_{,y} \rho_{,x,y}
			+ \rho_{,x}^2 \rho_{,y,y}}%
		{\kappa_\text{sqr}^2}\,,
	\\
	S &\coloneqq \frac{\rho_{,x}\rho_{,y}\rbr{\rho_{,x,x}-\rho_{,y,y}}
			-\rbr{\rho_{,x}^2-\rho_{,y}^2}\rho_{,x,y}}%
		{\kappa_\text{sqr}^2}\,.
	\label{eq:hr-dif-s}
\end{align}
\end{subequations}
Imposing $\kappa_\text{sqr}$ to be stationary in the direction of $w$ gives
\begin{align}
	0 = \frac{1}{2} \frpa{\kappa_\text{sqr}}{w} = \theta S\,, 
\end{align}
which gives either $\theta = 0$ or $S = 0$; they corresponds to the contour and
stream ridge-lines defined in eqs.\ \eqref{eq:dSVr-10} and \eqref{eq:stream-rid-lin-10}, 
respectively.

The contour and stream ridge-lines are distinct, except for two special cases.
Breton de Champ (see \cite[sec.\ 2]{rothe1915}) has shown that, 
\emph{stream-lines} satisfying $S = 0$ are 
\emph{necessarily straight lines}; otherwise, contour ridge-lines should not
be stream-lines, and they are therefore no stream ridge-line. However, it seems
to us that points satisfying $\rho_{,x} = \rho_{,y} =0$ also lie 
on both the contour and stream ridge-lines, see 
sections \ref{sec:asp-con} and \ref{sec:asp-str} for an example.

\begin{figure}
	\centering
	\begin{subfigure}{.49\linewidth}
	\begin{center}
		\includegraphics{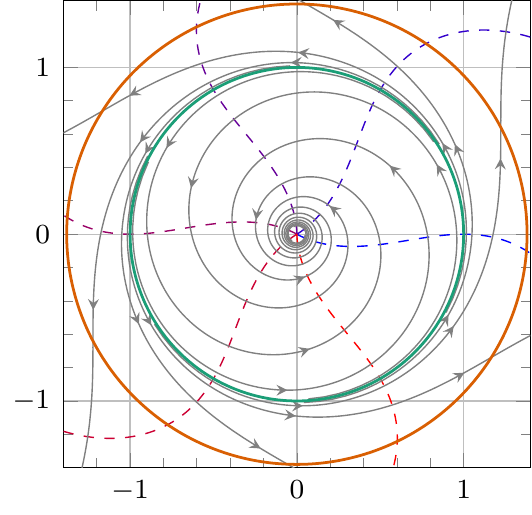}
	\end{center}
	\caption{Contour and stream ridge-lines}
	\label{fig:heli-g-10}
	\end{subfigure}
	\begin{subfigure}{.49\linewidth}
	\begin{center}
		\includegraphics{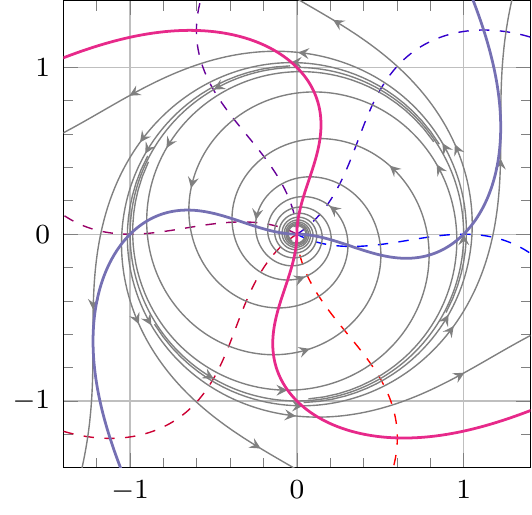}
	\end{center}
	\caption{First-derivative lines}
	\label{fig:heli-g-20}
	\end{subfigure}
	\begin{subfigure}{.49\linewidth}
	\begin{center}
		\includegraphics{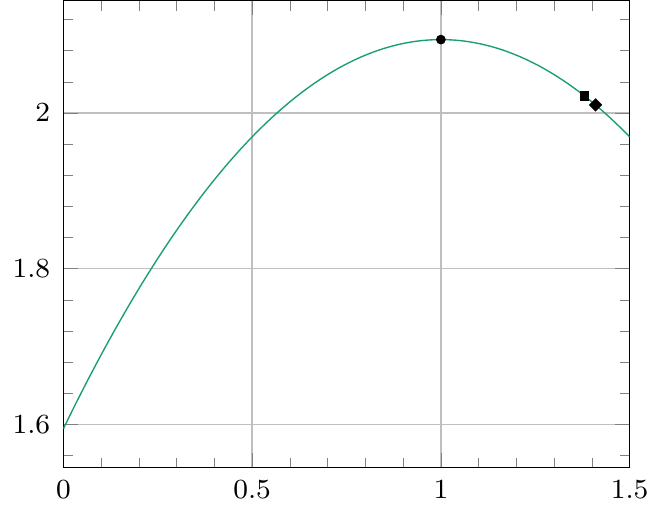}
	\end{center}
	\caption{Section at $\varphi = 2\pp/3$}
	\label{fig:heli-g-30}
	\end{subfigure}
	\caption[The helicoidal gutter and the ridge-lines]{%
	The so-called helicoidal gutter in eq.\ \eqref{eq:heli-100}, its contours (dashed
	lines) and gradient vector field (represented by the grey stream-lines
	with arrows), and its ridge-lines. In fig.\ \ref{fig:heli-g-10}, the green line 
	is the stream 	ridge-line given by $\theta = 0$, and the orange line is 
	the contour ridge-line predicted by the dSVr equation. In 
	fig.\ \ref{fig:heli-g-20}, the pink and the purple lines are $\rho_{,x} = 0$ and 
	$\rho_{,y} = 0$, 	respectively. In fig.\ \ref{fig:heli-g-30}, the section at 
	$\varphi = 2\pp/3$ is plotted, where the round, square and diamond points 
	are the stream and 	contour ridge-lines, as well as the first-derivative 
	line. One sees that it is the \emph{stream ridge-line} that picks the 
	highest point in the sense of constant $\varphi$-section.
	\label{fig:heli-10}}
\end{figure}

The differences, of the contour and stream ridge-lines, as well as the simple 
first-derivative test, can be shown with a so-called two-dimensional 
\emph{helicoidal gutter} \cite[sec.\ 6]{Koenderink1993}; in polar coordinates 
$\rbr{\varrho, \varphi}$ the metric and the gutter are
\begin{subequations}
\begin{align}
\begin{split}
	\dd s^2 &= g_{ij}\,\dd x^{i}\,\dd x^{j} =
	\dd \varrho^2 + \varrho^{2}\,\dd \varphi^2\,,
	\\
	\varrho &> 0\,, \quad 0 \le \varphi < 2\pp\,;
\end{split}
	\\
	\rfun{\rho}{\varrho, \varphi} &=
	\varphi + \frac{1}{2} \rbr{\frac{\varrho}{\varrho_0}-1}^2\,,
	\label{eq:heli-100}
\end{align}
\end{subequations}
see fig.\ \ref{fig:heli-10}. 

The contour ridge-lines of eq.\ \eqref{eq:heli-100} are given by the dSVr equation, 
or $S = 0$ in eq.\ \eqref{eq:hr-dif-s}. From the covariant expression in 
eq.\ \eqref{eq:dSVr-20}, one derives
\begin{align}
	0 = \rbr{\frac{\varrho}{\varrho_0}}^4 -
	\rbr{\frac{\varrho}{\varrho_0}}^3 - 1 \,.
\end{align}
The only positive root reads
\begin{align}
	\frac{\varrho}{\varrho_0} \approx \num{1.38028}\,.
\end{align}
See fig.\ \ref{fig:heli-g-10}. Roughly speaking, it crosses the contours where the 
latter are curved more.

As for the stream ridge-lines, using eq.\ \eqref{eq:stream-rid-lin-30} yields the 
equation for $w$
\begin{align}
	\dd \rho ^\sharp \intprod \dd w = 0\,,
	\qquad \text{or} \qquad
	0 = \rho_{,i} g^{ij} w_{,j} =
	\rbr{\frac{\varrho}{\varrho_0}-1} 
	\frac{w_{,\varrho}}{\varrho_0} + \frac{w_{,\varphi}}{\varrho^2}\,.
	\label{eq:heli-w-10}
\end{align}
The general integral to eq.\ \eqref{eq:heli-w-10} reads
\begin{align}
	w = \rfun{F}{-\frac{\varrho_0}{u} + \varphi -
		\rfun{\ln}{1-\frac{\varrho_0}{\varrho}}}\,,
\end{align}
where $F$ is an arbitrary function, see eq.\ \eqref{eq:theta-w-f}.
In order to obtain $\theta$, one applies eq.\ \eqref{eq:stream-rid-lin-05}
\begin{align}
	\theta\,\dd w = \star\dd \rho =
	-\frac{\rho_{,\varphi}}{\varrho}\,\dd \varrho +
	\varrho \rho_{,\varrho}\,\dd \varphi\,.
\end{align}
The result is
\begin{align}
	\theta = \frac{\varrho}{\varrho_0}
	\rbr{\frac{\varrho}{\varrho_0}-1} \cbr{\rfun{F'}{-\frac{\varrho_0}{\varrho}
			+\varphi - \rfun{\ln}{1-\frac{\varrho_0}{\varrho}}}}^{-1}\,.
\end{align}
The stream ridge-lines are then given by $\theta = 0$, or
\begin{align}
	\label{eq:stream-rl}
	\varrho = \varrho_0\,.
\end{align}
See fig.\ \ref{fig:heli-g-10}. One sees that the arbitrariness of $w$ encoded in
$F$ does \emph{not} affect the effectiveness of the stream approach.
Furthermore, the stream ridge-line really marks the
highest point for a constant $\varphi$-section. As a stream-line by itself, the 
stream ridge-line is also a limit cycle \cite[sec.\ 1.6.3]{ArnoldODEen} of the 
gradient vector field, and is also a watershed for two distinct families of
stream-lines, one spirals inwards and another outwards. The contour ridge-line,
on the other hand, is close to the highest point, see fig.\ \ref{fig:heli-g-20}.

Finally, the curves given by the first-derivative test  with respect to
$\rbr{x,y}$ can also be easily worked out, see fig.\ \ref{fig:heli-g-20}. They do 
not respect the rotational symmetry of $\rho$ and is therefore not very 
sensible. One may argue for an alternative test with respect to 
$\rbr{\varrho, \varphi}$, but the choice itself cannot be arbitrary and needs 
a mathematical description, which renders the method losing its simpleness.

%1234567890123456789012345678901234567890123456789012345678901234567890123456789
\section{Summary and outlook}
\label{sec:sum-out}

%123456789012345678901234567890123456789

%\subsection{Prospective applications}

%1234567890123456789012345678901234567890123456789012345678901234567890123456789

In current work, we have discussed the classical prediction from the ridge lines of stationary wave-packet in quantum theory. 
Our results show that the predictions from ridge lines are more abundant than the solutions solved from the classical theories; 
moreover, there may exist distinct deviation in the predictions from the classical solutions in certain range of minisupersapce, which arises from the quantum behaviour.  
This implies that the real classical trajectory should be corrected in this range.

First of all,
the stationary wave-packets are realised in quantum mechanics with the
superposition of degenerate energy eigenstates. Such cases are illustrated by
the toy model of a two-dimensional hydrogen atom. In reality, the Rydberg atom 
can also be described by such a superposition, providing a chance to verify the
theoretical statements.
Wave-packets constructed by superposing solutions of the Wheeler--DeWitt are 
also formally stationary. It is imaginable to make use of this fact and use
quantum systems in laboratory to simulate a quantum universe or a quantum
black hole.
However, one crucial difference between common quantum mechanical systems and
quantum cosmology is that, the latter usually has a \emph{Lorentzian} 
\enquote{kinetic energy term} in the Hamiltonian constraint, whereas the former
mostly have a \emph{Euclidean} kinetic energy term. One needs to be very 
creative to set up a simulated quantum cosmology system in laboratory.

%123456789

Secondly, the contour approach to ridge-lines, which dates back to Barr{\'e} de 
Saint-Venant in 1852, gives us an implicit equation \eqref{eq:dSVr-20} that can 
readily be plotted. It may not give results that are directionally minimal, but 
the difference can be small, see fig.\ \ref{fig:heli-g-30}.
The curves given by the dSVr equations are typically higher-order
algebraic equations, which can at least be numerically solved.
For the \enquote{linear} wave-packet, as well as for narrow Gaussian WKB 
wave-packets, this approach gives sensible results, as discussed
in section \ref{sec:hr-pg} and earlier in this section.
With a Euclidean signature, redundant curves can appear, as we have seen 
in this section with the modulated toy model $\rho = g\,\ee^{-f^2}$, as 
well as with the two-dimensional hydrogen atom, that may arise from the fact 
that the dSVr equations collect not only the most convex and concave 
neighbourhoods, but also the flattest points. For the toy model, the redundant
lines can be removed by careful asymptotic analysis, leaving results that also 
fit intuition.
As for the Lorentzian signature, however, it can happen that no result fully 
agrees with intuition, as we have seen in the modulated toy model. We have to 
decide whether to believe in mathematics and abandon our intuition, or stick to 
the intuition and find a better mathematical description.
Finally, an algorithm is needed to find the contour ridge-lines for numerically 
constructed wave-packets.
This is to be investigated in the future.

Lastly, the stream approach to ridge-lines, which dates back to Rudolf Rothe in 1915,
tells us to solve for a generic first integral $w$ of the gradient vector field
from eq.\ \eqref{eq:stream-rid-lin-07}, so that an inverse integral factor $\theta$
can be calculated, and $\theta = 0$ gives the singular stream-lines, that
define the stream ridge-lines. As has been shown with fig.\ \ref{fig:heli-g-30},
it can give results that are also directionally minimal.
With the Euclidean signature, directions of the gradient vector field give
the fastest ascent. The stream approach here gives results that
agree with intuitive expectations, and no redundant lines appear except for
those given by symmetries. We have shown this with the helicoidal gutter,
as well as the toy model $\rho=\ee^{-f^2}$ analytically; numerically, 
the modulated toy model $\rho = g\,\ee^{-f^2}$ as well as the two-dimensional
hydrogen atom also seem to perform pretty well under this approach.
As for the Lorentzian signature, the singular stream-lines of the gradient
vector field do not seem to agree with the intuitive ridge-lines, as we
have seen in the numeric results of the \enquote{linear} wave-packet in
fig.\ \ref{fig:2dhyd-sta-pac-30b}. The 
reason is that, for Lorentzian geometry, the directions of the gradient vector 
field differ from those of the gradient one-form field, and the former field 
no longer points to the direction of the fastest ascent. One can either discard
intuition and embrace what mathematical generalisation gives, or invent a 
novel notion of ridge-lines, keeping in mind that this new notion is also 
to work with the Euclidean case.
Finally, an algorithm is needed to find the singular ridge-lines for 
an analytically given gradient vector field, since the generic first integral
is difficult to solve. Moreover, for the cases where wave-packets are already
constructed numerically, another algorithm is needed to find the singular
ridge-lines from the numerically given gradient vector field.

The systematics of ridge-lines enables us to \emph{calculate} the classical
trajectories that emerge from a quantum wave-packet with \emph{arbitrary} width.
In fig.\ \ref{fig:linear-wp-rl-1der}, for example, one sees three trajectories, one of
which coincides or is close to a classical trajectory; with the profile of the
wave-packet considered, one may understand it as predicting a \emph{tunnelling} 
between two branches of the wave-packet, in that the wave-packet describes a
semi-classical universe evolving from one classical trajectory in the 
asymptotic region to \emph{another} classical trajectory, tunnelling near the 
origin of the plot. In contrast, the other two trajectories depart from 
classical trajectories near the classical turning point, giving a 
semi-classical behaviour that essentially differs from the classical one.
The tunnelling picture can be useful for the singularity avoidance, which also 
tells a semi-classical fate.

\section*{Acknowledgement}

The authors would like to thank Claus Kiefer (Cologne), Nick Kwidzinski (Düsseldorf), Leonardo Chataignier (Cologne), Tim Schmitz (Leverkusen), Ziping Rao (Bonn).

C.L.\ was supported by the Fundamental Research Funds for the Central Universities, Nankai University under the grant No.\ 63201006.
Y.-F.W.\ was supported by the \href{https://www.gradschool.physics.uni-koeln.de/}{Bonn-Cologne Graduate School of Physics and Astronomy (BCGS)}.

\appendix

%1234567890123456789012345678901234567890123456789012345678901234567890123456789
\section{Semi-classical approach of WKB}
\label{app:prototype-wkb}
%1234567890123456789012345678901234567890123456789012345678901234567890123456789

The WKB approach, named after Wentzel, Kramers and Brillouin
\cite{Wentzel1926,Kramers1926,Brillouin1926}, is an important approximation
in conventional quantum mechanics that separates the rapidly varying phase from 
the slowly varying amplitude [\citealp[ch.\ 7]{Landau3en}; 
\citealp[sec.\ 5.3.2]{Kiefer2012}]. It is also one of the
standard ways to connect quantum geometrodynamics with classical general 
relativity.

In contrast, the connection between the quantum and classical gravitational 
theories in the scenario of \emph{wave-packets}, is not very clear, and mostly 
\emph{ad hoc} case by case, shown with plots 
\cite{Kiefer1990,Andrianov2018wdx,Kiefer2019}. 
In \cite{Gerlach_1969}, the author observed that a superposition of WKB states 
can be chosen to have support only in a thin \enquote{tube} around a classical
trajectory. Moreover, in \cite{Hawking1986}, it was suggested that in the WKB 
approximation, an integral across a narrow section near a classical trajectory 
is related to the lapse function. Furthermore, in \cite{Lifschytz1994}, the 
author interpreted WKB wave-packets as containing higher-order WKB effects. And 
finally in \cite{Ohkuwa1996}, it was suggested that the wave function of the 
universe forms a narrow wave-packet in the classical region.

It is common to construct a wave-packet by superposing mode functions
with an amplitude that refers to a quantum number, e.g.\ superposing plane waves
with a Gaussian amplitude that refers to the momenta of the plane waves. At the
classical level, the quantum numbers correspond to first integrals, and using 
the former implies the existence of the latter. Therefore, this practice 
implicitly assumes that the system is \emph{Liouville integrable} 
\cite[sec.\ 49]{ArnoldMMCMen}, containing a number of first integrals. Systems
that do not have sufficient first integrals belong to the regime of classical
and quantum chaos \cite{Wimberger2014}, and will not be studied here. 
For a criterion of integrable systems that can be separated in the 
Hamilton--Jacobi formulation, see \cite{Waksjoe2003}.

In this section \ref{app:prototype-wkb}, we will first describe the general WKB 
theory in mathematics and minisuperspace models, and explain the
relation between the WKB mode functions and the classical trajectories
in section \ref{ssec:wkb-gt}. Then we will derive the WKB approximation 
for our prototype minisuperspace model, both by analysing the obtained
exact solution in section \ref{sec:asym-wkb}, and by working the WKB mode functions 
out from scratch in section \ref{sec:dirc-wkb}. Observing that these mode functions 
all contain a quantum number, we will show in section \ref{ssec:packet-phase} that
these quantum numbers have their correspondence at the classical level as
first integrals of the system, and the phase of the WKB mode functions is just 
the Hamilton's principal function. Finally, we will apply the theory 
established in section \ref{ssec:packet-phase} to wave-packets in 
section \ref{ssec:packet-gaussian}. We will show that these wave-packets, if 
constructed by superposing the WKB mode functions with a narrow Gaussian
amplitude, necessarily peak near a classical trajectory, which has the
first integrals corresponding to the centre of the Gaussian amplitude.

%1234567890123456789
\subsection{General theory}
\label{ssec:wkb-gt}
%1234567890123456789012345678901234567890123456789012345678901234567890123456789

This section \ref{ssec:wkb-gt} briefly introduces the WKB approximation in 
mathematics and the Wheeler--DeWitt approach.

Mathematically, the WKB approximation belongs to the class of \emph{global}
approximations to the solution of a linear differential equation, in which the
highest derivative is controlled by a small parameter $\delta$ 
\cite[ch.\ 10]{Bender1999}, with respect to which the solution 
$y = \rfun{y}{x}$ is expanded as a formal power series on the exponent:
\begin{align}
\rfun{\phi}{x} \sim \rfun{\exp}{\frac{1}{\delta}
	\sum_{n=0}^{+\infty} \delta^n \rfun{S_n}{x}}\,,
	\qquad \delta \to 0\,.
	\label{eq:prototype-wkb-sol-10}
\end{align}

In conventional quantum mechanics as well as in the Wheeler--DeWitt approach of 
quantum gravitation, the highest derivatives are controlled by the reduced
Planck constant $\hslash$. The meaning of a power expansion with respect to
such a \emph{dimensionful} quantity is questioned at the end of this subsection.

At the next-to-leading order, the WKB wave function is often taken as the
test solution \cite{Kiefer2019}
\begin{align}
	\psi \approx \sqrt{D}\,\ee^{\frac{\ii}{\hslash} S}\,,
	\label{eq:WKB-d-s-10}
\end{align}
where $S$ is the leading order term, $D = \ee^{\hslash^0 S_1}$ corresponds to
the real part of the next-to-leading order term,
which is called \emph{Van Vleck factor}, named after its eponymous
founder \cite{VanVleck1928}.\footnote{See \cite[ch.\ 7]{Pauli6de} for a viable
introduction of the Van Vleck factor; for historical remarks, see 
\cite{Visser1993,Wheeler2000_3}.} In the minisuperspace models, inserting 
eq.\ \eqref{eq:WKB-d-s-10} into the Wheeler-DeWitt equation, %(see e.g.\ \cite{Kiefer2012}), 
the resulting equations read \cite{Kiefer2019}
\begin{subequations}
\begin{align}
0 = \rfun{H_\perp}{\chi, \gamma; \frpa{S_0}{\chi}, \frpa{S_0}{\gamma}}
&= \frac{1}{2} \mscrG^{IJ} \frpa{S}{q^I} \frpa{S}{q^J}
	+ \mscrV(q)\,,
\label{eq:prototype-wkb-0-10} \\
\mscrG^{IJ} \frpa{S}{q^I} \frpa{D}{q^J} &=
- \rbr{\square S} D\,.
\label{eq:prototype-wkb-0-20}
\end{align}
\end{subequations}
Equation \eqref{eq:prototype-wkb-0-10} is just the \emph{Hamilton--Jacobi equation} for 
our singular system.
Results for the next orders can be found in e.g.\ 
\cite[sec.\ 5.4.1]{Kiefer2012}, which are not needed here.

%1234567890123456789
\subsection{Asymptotic expansion as a WKB approximation}
\label{sec:asym-wkb}
%1234567890123456789012345678901234567890123456789012345678901234567890123456789

In our prototype model, the exact solution of the minisuperspace 
Wheeler--DeWitt equation \eqref{eq:prototype-wdw-10} is known. The WKB approach
can therefore be realised in two ways. One can start with the generic WKB 
result, which means the Hamilton--Jacobi equation in 
\eqref{eq:prototype-wkb-0-10}, and then solve $S_0$ for it. This approach will 
be illustrated later in section \ref{sec:dirc-wkb}. Alternatively, one can also begin 
with the mode functions in eq.\ \eqref{eq:prototype-wdw-sol-10} which are exact 
solutions, and find an approximation for the Bessel functions that have the 
form of eq.\ \eqref{eq:prototype-wkb-sol-10}. We will follow this approach in this
section \ref{sec:asym-wkb}.

Since $\nu, x \propto \hslash^{-1}$ (c.f.\
eq.\ \eqref{eq:transform-20}), an approximation at \emph{small} $\hslash$ means
asymptotic expansion of the Bessel functions at \emph{large} $\nu$ \emph{and}
$x$. Note that
\begin{subequations}
\begin{align}
\rbr{\frac{\nu}{x}}^2 &=
	\frac{\varkappa p_{\gamma}^2}%
		{12\mathrm{Vol}_3^2 \vbr{V} \ee^{g \chi}}
	\\
&= \rfun{\mathrm{trig}}%
	{\sqrt{\frac{3}{2\varkappa}}g\rbr{\gamma - \gamma_0}}^2
	\qquad\text{by substituting eq.\ \eqref{eq:cla-traj-10}.}
\label{eq:nu-x-b}
\end{align}
\end{subequations}
Equation \eqref{eq:nu-x-b} makes sense \emph{if} we want to study the behaviour of the 
mode functions \emph{near} a classical trajectory.

In such a case of fixed $\nu / x$, the asymptotic representations belong to the 
\enquote{Debye} type \cite[sec.\ 3.14.2]{Magnus1966}. In the following we give
the leading order results. For the $\rbr{-, +}$ case with $\rfun{J_\nu}{x}$, 
the Debye expansion reads \cite[eq.\ (10.19.6)]{NIST:DLMF}
\begin{align}
\begin{split}
\rfun{J_{\nu}}{x} &= \sqrt{\frac{2}{\pp}}\rbr{x^2-\nu^2}^{-1/4}
	\\
	&\quad\,\cdot
	\cbr{\sfun{\sin}{\sqrt{x^2-\nu^2}-\nu \arccos\frac{\nu}{x} + \frac{\pp}{4}}
	+ \rfun{\Omicron}{x^{-1}}}
	\qquad x > \nu\,,
\end{split}
\label{eq:debye-J-10}
\end{align}
where $x > \nu$ holds because $\mathrm{trig} = \sin$ for $\rbr{-, +}$, and 
$x = \nu$ is excluded because it is not contained in the trajectories.
%, see fig.\ \ref{fig:cla-traj-10--+}.
The mode function 
$\ee^{\frac{\ii}{\hslash} p_{\gamma} \rbr{\gamma - \tilde{\gamma}_0}}
 \rfun{J_{\nu}}{x}$
contains therefore \emph{two} WKB branches 
$\sim\ee^{\frac{\ii}{\hslash} S_{\pm}}$,
\begin{align}
\frac{S_{\pm}}{\hslash} \coloneqq \frac{p_{\gamma}}{\hslash} 
\rbr{\gamma - \tilde{\gamma}_0}
	\pm \rbr{\sqrt{x^2-\nu^2}-\nu \arccos\frac{\nu}{x} + \frac{\pp}{4}}\,.
	\label{eq:wkb-J-10}
\end{align}
Note we have introduced an additive
constant $\tilde{\gamma}_0$ to cancel the extra constant factors and match the
classical constant $\gamma_0$, which is also related to
eqs.\ \eqref{eq:prin-cons-int-10} and
\eqref{eq:prin-cons-int-20}. By using
eq.\ \eqref{eq:prin-cons-int-10}, one gets
\begin{align}
0 = \frac{1}{\hslash} \frpa{S_{\pm}}{p_{\gamma}}
= \rbr{\gamma-\gamma_0} \mp \sqrt{\frac{2\varkappa}{3}}\,\frac{1}{g}
\arccos \sqrt{\frac{\varkappa p_{\gamma}^2}%
	{12\mathrm{Vol}_3^2 \vbr{V}\ee^{g\chi}}}\,,
\end{align}
which leads to eq.\ \eqref{eq:cla-traj-10} with $\mathrm{trig} = \sin$. 

For the $\rbr{+, -}$ case with $\rfun{F_{\ii \nu}}{x}$ and 
$\rfun{G_{\ii \nu}}{x}$, the Debye expansions at leading order read
\cite[eqs.\ (5.15) and (5.16)]{dunster1990}
\begin{subequations}
\begin{align}
\begin{split}
\rfun{F_{\ii \nu}}{x} &= \sqrt{\frac{2}{\pp}}\rbr{x^2+\nu^2}^{-1/4}
	\\ &\quad\,\cdot
	\cbr{\sfun{\sin}{\sqrt{x^2+\nu^2}-\nu \arcsinh\frac{\nu}{x} + \frac{\pp}{4}}
	+ \rfun{\Omicron}{x^{-1}}}\,,
\end{split}
\label{eq:debye-F-10} \\
\begin{split}
\rfun{G_{\ii \nu}}{x} &=-\sqrt{\frac{2}{\pp}}\rbr{x^2+\nu^2}^{-1/4}
	\\ &\quad\,\cdot
	\cbr{\sfun{\cos}{\sqrt{x^2+\nu^2}-\nu \arcsinh\frac{\nu}{x} + \frac{\pp}{4}}
	+ \rfun{\Omicron}{x^{-1}}}\,,
\end{split}
\label{eq:debye-G-10}
\end{align}
\end{subequations}
where $x, \nu \in \BbbR^+$ are arbitrary. Both cases contain two WKB branches. 
Take $\rfun{F_{\ii \nu}}{x}$ as an example, one has
\begin{align}
\frac{S_{\pm}}{\hslash} &= \frac{p_{\gamma}}{\hslash} 
\rbr{\gamma - \tilde{\gamma}_0}
	\pm \rbr{\sqrt{x^2+\nu^2}-\nu \arcsinh\frac{\nu}{x} + \frac{\pp}{4}}\,,
	\label{eq:wkb-F-10}\\
0 &= \frac{1}{\hslash} \frpa{S_\pm}{p_\gamma} =
\rbr{\gamma-\gamma_0} \mp \sqrt{\frac{2\varkappa}{3}}\,\frac{1}{g}
\arcsinh \sqrt{\frac{\varkappa p_{\gamma}^2}%
	{12\mathrm{Vol}_3^2 \vbr{V}\ee^{g\chi}}}\,,
\end{align}
which also leads to eq.\ \eqref{eq:cla-traj-10} with $\mathrm{trig} = \sinh$. The 
calculation for $\rfun{G_{\ii \nu}}{x}$ is essentially the same, with an extra
constant phase shift $\pp/2$.

Finally, for the $\rbr{+, +}$ case, the expansion at leading order reads 
\cite[p.\ 141--142]{Magnus1966}
\begin{align}
\begin{split}
\rfun{K_{\ii \nu}}{x} &=
	\sqrt{\frac{2\pp}{\ee^{\pp \nu}}} \rbr{\nu^2 - x^2}^{-1/4}
	\\ & \quad\,\cdot
	\cbr{\sfun{\cos}{\sqrt{\nu^2 - x^2} - \nu \arccosh \frac{\nu}{x}
			+ \frac{\pi}{4}} + \rfun{\Omicron}{x^{-1}}}
	\qquad \nu > x\,,
	\end{split}
	\label{eq:debye-K-10}
\end{align}
where $\nu > x$ holds because $\mathrm{trig} = \cosh$ for $\rbr{+, +}$.
Equation \eqref{eq:debye-K-10} contains, once again, two WKB branches, and one has
\begin{align}
\frac{S_{\pm}}{\hslash} &= \frac{p_{\gamma}}{\hslash} 
\rbr{\gamma - \tilde{\gamma}_0}
	\pm \rbr{\sqrt{\nu^2-x^2} - \nu \arccosh \frac{\nu}{x} + \frac{\pi}{4}}\,,
	\label{eq:wkb-K-10} \\
0 &= \frac{1}{\hslash} \frpa{S_{\pm}}{p_{\gamma}} = 
\rbr{\gamma-\gamma_0} \mp \sqrt{\frac{2\varkappa}{3}}\,\frac{1}{g}
\arccosh \sqrt{\frac{\varkappa p_{\gamma}^2}%
	{12\mathrm{Vol}_3^2 \vbr{V}\ee^{g\chi}}}\,,
\end{align}
which, again, lead to eq.\ \eqref{eq:cla-traj-10} with $\mathrm{trig} = \cosh$.

%1234567890123456789
\subsection{WKB approximation by direct calculation}
\label{sec:dirc-wkb}
%1234567890123456789012345678901234567890123456789012345678901234567890123456789

In this section \ref{sec:dirc-wkb}, we obtain the WKB phase $S$ and the van Vleck
factor $D$ directly from eqs.\ \eqref{eq:prototype-wkb-0-10} and \eqref{eq:prototype-wkb-0-20}. 

To begin with, one can verify that the $S_{\pm}$ given by 
eqs.\ \eqref{eq:wkb-J-10}, \eqref{eq:wkb-F-10} and \eqref{eq:wkb-K-10} are indeed complete integrals of the
Hamilton--Jacobi equation \eqref{eq:prototype-wkb-0-10}, which is a 
\emph{non-linear} first-order partial differential equation by itself.

The transport equation \eqref{eq:prototype-wkb-0-20} in our prototype model
reads
\begin{align}
\mitsanss \frac{\hslash}{\mathrm{Vol}_3}
	\rbr{-\frac{\varkappa}{6} \frpa{S}{\gamma}\frpa{D}{\gamma}
		+ \mitsansl \frpa{S}{\chi}\frpa{D}{\chi} }
	&= -\frac{\mitsanss}{\mathrm{Vol}_3}\rbr{
			- \frac{\varkappa}{6} \frpa{^2 S}{\gamma^2} +
			\mitsansl \frpa{^2S}{\chi^2}} D\,,
	\label{eq:van-vleck-10}
\end{align}
which is a first-order \emph{linear} partial differential equation. By using 
the transformation in eq.\ \eqref{eq:transform-20}, we are able to derive the 
\emph{general} integral, which contains an arbitrary \emph{function} $D_0$, in 
contrary to the \emph{complete} integrals for $S$, where merely arbitrary 
\emph{constants} are present. See table \ref{tab:van-vleck-10}.
\begin{table}
	\begin{center}
	\begin{tabular}{l|c}
	\toprule
	$\rbr{\mitsansl, \mitsanss \mitsansv}$ & $D_\pm$ \\
	\midrule
	$\rbr{-, -}$ & no solution \\
	$\rbr{-, +}$ & $\rbr{+x^2-\nu^2}^{-1/2}
		\rfun{D_0}{\sqrt{\frac{3}{2\varkappa}}\,g \gamma
			\mp \arccos\frac{\nu}{x}}$\\
	$\rbr{+, -}$ & $\rbr{+x^2+\nu^2}^{-1/2}
		\rfun{D_0}{\sqrt{\frac{3}{2\varkappa}}\,g \gamma
			\mp \arcsinh\frac{\nu}{x}}$\\
	$\rbr{+, +}$ & $\rbr{-x^2+\nu^2}^{-1/2}
		\rfun{D_0}{\sqrt{\frac{3}{2\varkappa}}\,g \gamma
			\mp \arccosh\frac{\nu}{x}}$\\
	\bottomrule
	\end{tabular}
	\end{center}
\caption[General integrals of the Van Vleck factor $D_{\pm}$]%
	{General integrals of the Van Vleck factor $D_{\pm}$ that
	are solutions to eq.\ \eqref{eq:van-vleck-10} and correspond to $S_{\pm}$.
	The pre-factors are in accordance with those in 
	eqs.\ \eqref{eq:debye-J-10}, \eqref{eq:debye-F-10},
	\eqref{eq:debye-G-10} and \eqref{eq:debye-K-10}.
	The arbitrary function $D_0$ can be argued to be a constant.}
\label{tab:van-vleck-10}
\end{table}

Since $S_{\pm}$'s are complete integrals that result from separation test
solutions (see eq.\ \eqref{eq:packet-250} below), the full Van Vlack factor should
also be in a separated form, which would render $D_0$ constant, because
it mixes $\gamma$ with $\chi$ otherwise. This can be verified if one begins 
from scratch by inserting the WKB wave function in eq.\ \eqref{eq:WKB-d-s-10} into 
the Wheeler--DeWitt equation \eqref{eq:prototype-wdw-10}, and then adapts a 
separation test solution. An ordinary differential equation in $\chi$
would arise, from which one could find the second terms of $S_{\pm}$'s in 
eqs.\ \eqref{eq:debye-J-10}, \eqref{eq:debye-F-10},
\eqref{eq:debye-G-10} and
\eqref{eq:debye-K-10}
that only contain $\gamma$, and the corresponding $D_{\pm}$'s are solved
by the pre-factors in table \ref{tab:van-vleck-10}, with no place for the
arbitrary function $D_0$.

We conclude that eqs.\ \eqref{eq:prototype-wkb-0-20} and \eqref{eq:van-vleck-10} may not be the 
best starting point to solve for the Van Vleck factor for systems with multiple
degrees of freedom.

%123456789012345678901234567890123456789
\subsection{WKB phase as a complete integral}
\label{ssec:packet-phase}
%1234567890123456789012345678901234567890123456789012345678901234567890123456789

In this section \ref{ssec:packet-phase}, we study the WKB mode functions and their 
phases. We will see that the mode functions can be chosen, such that they are 
labelled with quantum numbers, which are related to classical integrals of
motion. Correspondingly, their phases are complete integrals of the classical
Hamilton--Jacobi equation, which contain the classical integrals mentions 
above.

For the Hamilton--Jacobi equation \eqref{eq:prototype-wkb-0-10}, the useful
family of solutions is the \emph{complete solution} or \emph{complete integral}
[\citealp[sec.\ 47]{Landau1en}; 
\citealp[sec.\ 3.1]{evans2010partial};
\citealp[sec.\ 9.4]{ArnoldMMCMen}],
that containing integral constants, e.g.
\begin{align}
	S = \rfun{S}{q^i; \alpha_1, \ldots, \alpha_{n-1}} + \alpha_n\,,
	\label{eq:packet-150}
\end{align}
where $\alpha_i$ are constants, $i = 1, 2, \ldots, n$.
A classical trajectory that corresponds to 
this WKB solution can then be obtained by the
\emph{principle of constructive interference} \cite{Gerlach_1969} as
\begin{subequations}
\begin{align}
	\frpa{S}{\alpha_i} = 0\,.
	\label{eq:prin-cons-int-10}
\end{align}
Meanwhile, in the classical Hamilton--Jacobi formalism, the related equations 
are
\begin{align}
	\frpa{S}{\alpha_i} = \beta_i\,,
	\label{eq:prin-cons-int-20}
\end{align}
\end{subequations}
where $\cbr{\alpha_i}$'s are the constants contained in the complete integral 
$S$, and $\cbr{\beta_i}$'s are another set of constants 
\cite[sec.\ 47]{Landau1en}. 

Now, if $S$ is a complete integral in the form of eq.\ \eqref{eq:packet-150}, a 
stationary wave-packet can be constructed by smearing out each constant with an 
amplitude, see e.g.\ section \ref{ssec:packet-gaussian}. 

In practice, it has been shown in \cite{Gerlach_1969} that, in order to be 
able to derive the Hamilton equations for the canonical momenta in full
geometrodynamics, it is sufficient and necessary that $S$ is a complete 
integral of the Hamilton--Jacobi equation, containing a number of constants 
that is equal to the physical degrees of freedom.

In the following we give a construction, in which the phase factor $S$ in
eq.\ \eqref{eq:WKB-d-s-10} is indeed of a form close to the expression in
eq.\ \eqref{eq:packet-150}. Let the system be such that $m \le n-1$ variables can be
iteratively separated \cite[sec.\ 48]{Landau1en}, so that the following
equations can be obtained along a classical trajectory
\begin{align}
\begin{split}
	\rfun{\phi_1}{q^1, \frde{S_1}{q^1}} \eqqcolon \alpha_1\,,
	\qquad
	\rfun{\phi_2}{q^2, \frde{S_2}{q^2}; \alpha_1} &\eqqcolon \alpha_2\,,
	\ldots\,,\\
	\rfun{\phi_m}{q^m, \frde{S_n}{q^m}; \alpha_1,\ldots, \alpha_{m-1}} 
	&\eqqcolon \alpha_m\,,
\end{split}
\label{eq:packet-200}
\end{align}
and the corresponding complete integral, \eqref{eq:packet-150}, reads
\begin{align}
\begin{split}
	\rfun{S}{q^1, \ldots, q^n; \alpha_1, \ldots \alpha_m} &=
	\rfun{S_1}{q^1; \alpha_1} + \ldots + 
	\rfun{S_m}{q^m; \alpha_1, \ldots, \alpha_m}
	\\
	&\quad\,+
	\rfun{S_{m+1}}{q^{m+1}\ldots q^n; \alpha_1, \ldots, \alpha_m}\,.
	\label{eq:packet-250}
\end{split}
\end{align}
From the Hamilton--Jacobi theory, we know that 
$\cbr{\rfun{\phi_j}{q^j, p_j}}$'s are \emph{in involution} 
\cite[sec.\ 10.1]{ArnoldMMCMen} with $H_\perp$, 
i.e.\ the Poisson brackets vanish,
\begin{align}
	\sbr{\rfun{\phi_j}{q^j, p_j},
		\rfun{H_\perp}{q^1, \ldots, q^n, p_1, \ldots, p_n}}_\text{P} = 0\,,
	\qquad \forall j = 1, \ldots, m\,.
\end{align}
Furthermore, we require that $\cbr{\rfun{\phi_j}{q^j, p_j}}$'s are in 
\emph{mutual} involution.

Upon canonical quantisation, the $H_\perp$ and $\cbr{\phi_j}$'s are promoted to
(if necessary, \emph{self-adjoint}) %, see section \ref{sec:sa})
operators 
\cite[sec.\ 5.1]{Kiefer2012}, and the condition of mutual involution with
respect to $\sbr{\cdot, \cdot}_{\text{P}}$ is promoted to commuting
$\frac{1}{\ii \hslash} \sbr{\cdot, \cdot}_{-}$. Equation \eqref{eq:packet-200} are 
promoted to the simultaneous eigenvalue equations
\begin{align}
\begin{split}
	\rfun{\phi_1}{q^1, \frac{\hslash}{\ii}\partial_1} \psi
		= \alpha_1 \psi\,,
	\qquad
	\rfun{\phi_2}{q^2, \frac{\hslash}{\ii}\partial_2; \alpha_1} \psi &=
		\alpha_2 \psi\,,
\\
	\ldots\,,\qquad
	\rfun{\phi_n}{q^n, \frac{\hslash}{\ii}\partial_m; \alpha_1,
		\ldots, \alpha_{m-1}}\psi &= \alpha_m \psi\,,
\end{split}
\label{eq:packet-300}
\end{align}
so that one can write $\psi = \psi_{\alpha_1\ldots\alpha_m}$. Applying a WKB 
test solution to eq.\ \eqref{eq:packet-300} results in the WKB wave function in 
eq.\ \eqref{eq:WKB-d-s-10} with $S$ given by eq.\ \eqref{eq:packet-250}. This finishes our
construction.

\bibliographystyle{JHEP}
\bibliography{main.bib}

\end{document}

%% file: preambles/preambles.tex
\usepackage{attrib}

\usepackage{amssymb}

\usepackage{mathtools}
\usepackage{braket}
\usepackage{siunitx}
\usepackage{graphicx}
\usepackage{empheq}
\usepackage{amsthm}

\usepackage{enumitem} % \begin{enumerate}[label=\Alph*]

\usepackage{multirow}
\usepackage{booktabs}
\usepackage{cancel}
\usepackage{nameref}
\usepackage{bookmark}

\usepackage{subcaption}

\usepackage{rotating}

\usepackage{tikz}

\usepackage{pgfplots}
\usepgfplotslibrary{fillbetween,colorbrewer}
\usetikzlibrary{external,
	angles,
	decorations.pathmorphing,
	decorations.markings,
	calc,intersections,through,backgrounds,
	decorations.pathreplacing,
	patterns,shadings,
	arrows,
	shapes.geometric}

\usepackage{pdftexcmds}
\makeatletter % if not in a .sty file
\newcommand{\includetikz}[1]{%
	\ifnum\pdf@shellescape=1
	\tikzsetnextfilename{#1}
	\input{./tikz/#1.tex}
	\else
	\includegraphics{./figures/#1.pdf}
	\fi
}
\makeatother

\pgfplotsset{
	samples=128,
	cycle list/Dark2-8,
	cycle multiindex* list={
		mark list*\nextlist
		Dark2-8\nextlist
	},
	every axis/.append style={
		minor tick num=4,
		grid=major,
		width=.99\textwidth,
		label style={font=\scriptsize},
		tick label style={font=\scriptsize},
	},
	legend style={
		legend columns=2,
		fill opacity=.6,
		draw opacity=1,
		text opacity=1,
		font=\scriptsize,
	},
	/pgfplots/layers/Bowpark/.define layer set={
        axis background,axis grid,main,axis ticks,axis lines,axis tick labels,
        axis descriptions,axis foreground
    }{/pgfplots/layers/standard},
	compat=newest,
}
\pgfmathsetmacro\kreiszahl{pi}
\pgfmathsetmacro\nSample{128}

\usepackage{csquotes}

\usepackage{physics}

%% file: preambles/math.tex
\usepackage{upgreek}

\newcommand{\ii}{\mathsf{i}}
\newcommand{\ee}{\mathsf{e}}
\newcommand{\pp}{\uppi}

\newcommand{\intprod}{\lrcorner}
\newcommand{\Omicron}{O}
\newcommand{\mitsansg}{\mathsf{g}}
\newcommand{\mitsansl}{\mathsf{l}}
\newcommand{\mitsanss}{\mathsf{s}}
\newcommand{\mitsansv}{\mathsf{v}}

\newcommand{\mscrA}{\mathcal{A}}
\newcommand{\mscrG}{\mathcal{G}}
\newcommand{\mscrV}{\mathcal{V}}

\newcommand{\BbbN}{\mathbb{N}}
\newcommand{\BbbR}{\mathbb{R}}

\newcommand{\Alpha}{A}
\newcommand{\Rho}{P}

\DeclareMathOperator{\arcsinh}{arsinh}
\DeclareMathOperator{\arccosh}{arcosh}

\DeclareMathOperator{\sgn}{sgn}
\usepackage{stackengine}
\stackMath
\newcommand\tenq[2][1]{%
 \def\useanchorwidth{T}%
  \ifnum#1>1%
    \stackunder[0pt]{\tenq[\numexpr#1-1\relax]{#2}}{\scriptscriptstyle\sim}%
  \else%
    \stackunder[1pt]{#2}{\scriptscriptstyle\sim}%
  \fi%
}

\stackMath
\newcommand\tsup[2][2]{%
 \def\useanchorwidth{T}%
  \ifnum#1>1%
    \stackon[-.5pt]{\tsup[\numexpr#1-1\relax]{#2}}{\scriptscriptstyle\sim}%
  \else%
    \stackon[.5pt]{#2}{\scriptscriptstyle\sim}%
  \fi%
}

\usepackage{graphicx,stackengine,scalerel}

\newcommand{\rbr}[1]{{\left(#1\right)}}
\newcommand{\sbr}[1]{{\left[#1\right]}}
\newcommand{\cbr}[1]{{\left\{#1\right\}}}
\newcommand{\abr}[1]{{\left<#1\right>}}
\newcommand{\vbr}[1]{{\left|#1\right|}}

\newcommand{\rfun}[2]{{#1}\mathopen{}\left(#2\right)\mathclose{}}
\newcommand{\sfun}[2]{{#1}\mathopen{}\left[#2\right]\mathclose{}}
\newcommand{\cfun}[2]{{#1}\mathopen{}\left\{#2\right\}\mathclose{}}

\newcommand{\frde}[2]{{\frac{\dd{#1}}{\dd{#2}}}}

\newcommand{\frpa}[2]{{\frac{\partial{#1}}{\partial{#2}}}}